\documentclass{jfm}
\usepackage{graphicx}
\usepackage{epstopdf, epsfig}
\usepackage{newtxtext}
\usepackage{newtxmath}
\usepackage{natbib}

\shorttitle{Effect of electrostatic interaction}
\shortauthor{X. Ruan, M. Gorman and R. Ni}

\title{Effects of electrostatic interaction on clustering and collision of bidispersed inertial particles in homogeneous and isotropic turbulence}

\author{Xuan Ruan,
  Matthew T. Gorman
 \and Rui Ni \corresp{\email{rui.ni@jhu.edu}}}

\affiliation{Department of Mechanical Engineering, Johns Hopkins University, Baltimore, MD 21218, USA}

\begin{document}

\maketitle

\begin{abstract}
In sandstorms and thunderclouds, turbulence-induced collisions between solid particles and ice crystals lead to inevitable triboelectrification. The charge segregation is usually size-dependent, with small particles charged negatively and large particles charged positively. In this work, we perform numerical simulations to study the influence of charge segregation on the dynamics of bidispersed inertial particles in turbulence. Direct numerical simulations of homogeneous isotropic turbulence are performed with the Taylor Reynolds number $\mathrm{Re}_{\lambda}=147.5$, while particles are subjected to both electrostatic interactions and fluid drag, with Stokes number of 1 and 10 for small and large particles, respectively. Coulomb repulsion/attraction are shown to effectively inhibit/enhance particle clustering within a short range. Besides, the mean relative velocity between same-size particles is found to rise as the particle charge increases because of the exclusion of low-velocity pairs, while the relative velocity between different-size particles is almost unaffected, emphasizing the dominant roles of differential inertia. The mean Coulomb-turbulence parameter, $\mathrm{Ct}_0$, is then defined to characterize the competition between the Coulomb potential energy and the mean relative kinetic energy. In addition, a model is proposed to quantify the rate at which charged particles approach each other and captures the transition of the particle relative motion from the turbulence-dominated regime to the electrostatic-dominated regime. Finally, the probability distribution function of the approaching rate between particle pairs are examined, and its dependence on the Coulomb force is further discussed using the extended Coulomb-turbulence parameter.
\end{abstract}


\section{Introduction}
Particle-laden turbulent flows are widespread in both nature and industry. One of the most important features of turbulence is its ability to effectively transport suspended particles and create highly inhomogeneous particle distributions, which eventually lead to frequent particle collisions or bubble/droplet coalescence \citep{FalkovichRMP2001, ToschiARFM2009, PumirARFM2016}. Typical natural processes can be found as rain formation \citep{ShawARFM2003,GrabowskiARFM2013}, sandstorms \citep{ZhangNatComm2020}, and the formation of marine snow in the ocean \citep{ArguedasPNAS2022}, while in industrial applications turbulence-induced collisions are also shown essential in dusty particle removal \citep{JaworekPECS2018} and flocculation \citep{ZhaoJFM2021}.

It has been widely shown that the spatial distribution of inertial particles is highly nonuniform and intermittent, which is known as the preferential concentration \citep{MaxeyJFM1987, SquiresPOF1991}. Preferential concentration could significantly enhance local particle number density and facilitate interparticle collisions \citep{ReadePOF2000}. At the dissipative range, clustering is driven by Kolmogorov-scale eddies that eject heavy particles out of high-vorticity regions while entrap light bubbles \citep{BalkovskyPRL2001, CalzavariniPRL2008, MathaiPRL2016}. Moreover, later studies show that in addition to clustering at the dissipative scale, the coexistence of multi-scale vortices in turbulence also affect the dynamics of heavy particles with relaxation times much larger than the Kolmogorov time scale, and drive self-similar clustering at larger scales \citep{BecPRL2007,GotoPOF2006, YoshimotoJFM2007, BakerJFM2017}. Besides, the inertial-range clustering of heavy particles can be alternatively explained by the sweep-stick mechanism, where inertial particles are assumed to stick to and move with fluid elements with zero acceleration \citep{GotoPRL2008}.

Apart from preferential concentration, the sling effect or the formation of caustics becomes more significant with the growth of particles inertia or turbulence intensity \citep{FalkovichNature2002, WilkinsonEuroPhysLett2005, WilkinsonPRL2006, FalkovichJAtomSci2007}. When sling happens, inertial particles encountering intermittent fluctuations get ejected out of local flow. Then clouds of particles with different flow histories interpenetrate each other with large relative velocities \citep{BewleyNJP2013}. As a result, the large relative velocity gives rise to an abrupt growth of the collision frequency, and the sling effect becomes the dominant collision mechanism for large-inertia particles \citep{VosskuhleJFM2014}.

By means of direct numerical simulations (DNS), the contributions of both preferential concentration and the sling effect (or the turbulence transport effect) to the collision rate could be quantified separately. A framework based on the radial distribution function and the mean relative velocity was proposed to predict the collision kernel function of neutral monodisperse particles with arbitrary inertia \citep{SundaramJFM1997, WangJFM2000}. Based on this framework, the impacts of turbulence Reynolds number, nonlinear drag, gravity, and hydrodynamic interactions were systematically investigated \citep{AyalaNJP2008a, AyalaNJP2008b, IrelandJFM2016a, IrelandJFM2016b, AbabaeiJFM2021, BraggJFM2022}. In addition to monodisperse particles, this framework was also extended to bidispersed particle systems. Compared to monodisperse system, the clustering of different-size particles is found weaker, while the turbulent transport effect is more pronounced due to particles' different responses to turbulent fluctuations \citep{ZhouJFM2001}.

In addition to neutral particles, particles in nature could easily get charged after getting through ionized air \citep{MarshallCambridge2014, VanMinderhoutNatComm2021} or encountering collisions \citep{LeeNatPhys2015}. The resulting electrostatic interaction could significantly modulate the fundamental particle dynamics in turbulence, such as clustering, dispersion and agglomeration \citep{KarnikPOF2012, YaoPRF2018, RuanJFM2021, BoutsikakisJFM2022, RuanPRE2022, ZhangJFM2023}. For identically charged particles, it has been shown that the Coulomb repulsion could effectively repel close particle pairs and mitigate particle clustering. By assuming that the drift velocity caused by Coulomb force can be superposed to the turbulence drift velocity, the radial distribution function of particles with negligible/finite inertia can be derived, which shows good agreement with both experimental and numerical results \citep{LuPRL2010, LuNJP2010, LuPOF2015}. 

Different from the like-charged monodisperse system, the real charged particle-laden flows may be more complicated because: (1) the size of suspended particles is generally polydispersed \citep{ZhangNatComm2020}, and (2) the charge polarity varies with the particle size because of charge segregation. That is, after triboelectrification, smaller particles tend to accumulate negative charges, while large ones are preferentially positively charged \citep{Forward2009, LeeNatPhys2015}. In the previous work by \citet{DiRenzo2018}, DNS of the bidispersed suspensions of oppositely charged particles was performed. It was shown that, the aerodynamic responses of different-size particles to turbulent fluctuations varies significantly. Although the system is overall neutral, different clustering behaviors lead to a spatial separation of positive and negative charge generating mesoscale electric fields in the space. 

However, even though the charge segregation is often correlated with the size difference, few studies have been devoted to the dynamics of bidispersed particles with both charge and size difference. In this system, three kinds of electrostatic interactions are introduced simultaneously, i.e., repulsion between small-small/large-large particles and attraction between small-large particles. How would these electrostatic forces affects the clustering of bidispersed particles is less understood. Moreover, the collision frequency of charged particles is of great importance to understand the physics of rain initiation \citep{HarrisonPRL2020} and planet formation \citep{SteinpilzNatPhys2020}, and the influence of charge segregation on these processes are still unclear.

To address the above issues, in this study we perform direct number simulations to investigate the dynamics of bidispersed oppositely charged particles in homogeneous isotropic turbulence. The numerical methods are introduced in Sec.\ref{sec:Methods}. In Sec.\ref{sec:Results_and_Discussions}, the statistics of the radial distribution function and the mean relative velocity are first presented to show the effects of electrostatic force on the clustering and the relative motion between different kinds of particle pairs. Then the modulation of collision frequency is quantified using the particle flux, and the mean Coulomb-turbulence parameter is proposed to characterize the relative importance of electrostatic force compared to particle-turbulence interaction. Finally, the particle flux density in the relative velocity space is discussed, which illustrates in detail how electrostatic force affects the collision of charged particle pairs with different relative velocities.
 
\section{Methods}
\label{sec:Methods}
\subsection{Simulation conditions}

The transport of bidispersed solid particles in turbulence is numerically investigated using the Eulerian-Lagrangian framework. The flow field is homogeneous isotropic turbulence with $\mathrm{Re}_{\lambda}=147.5$. Solid particles are treated as discrete points and their motions are updated based on Maxey-Riley equation \citep{MaxeyPOF1983}. 

Table \ref{tab:SimulationParameters} lists simulation parameters used in this study. Particle properties are chosen based on typical values of silica particles suspended in gaseous turbulence. Here, the particle density is much larger than that of the air ($\rho_{\mathrm{p}}/\rho_{\mathrm{f}} \sim O(10^3)$), and the particle size is smaller than the Kolmogorov length (i.e.,  $d_{\mathrm{p,S}}/\eta=0.1, d_{\mathrm{p,L}}/\eta \approx 0.3$). The particle Stokes $St$, defined as the ratio of the particle relaxation time $\tau_{\mathrm{p}}$ to the Kolmogorov time scale $\tau_{\eta}$, is given by

\begin{equation}
    St = \frac{\tau_{\mathrm{p}}}{\tau_{\eta}} = \frac{1}{18} \frac{\rho_{\mathrm{p}}}{\rho_{\mathrm{f}}} {\left( \frac{d_{\mathrm{p}}}{\eta} \right)}^2.
\end{equation}

\noindent
In previous field measurement \citep{ZhangNatComm2020}, the Stokes number of sand particles mainly lies with the range of $St = 1 - 10$ in strong sandstorms. We thus choose $St_{\mathrm{S}}=1$ and $ St_{\mathrm{L}}=10$ as the typical Stokes numbers for small and large particles, respectively. In addition, since the measured particle mass concentration in \citet{ZhangNatComm2020} is low, the particle volume fraction is set small ($\phi_{\mathrm{par}} = 6.87 \times 10^{-6}$) in our simulation, and the force feedback from the dilute particle phase to the fluid phase is omitted \citep{BalachandarARFM2010}.

When choosing the particle charge, we assume that particles carry a certain amount of charge as a result of triboelectrification. Both small and large particles are assigned the same amount of charge $q$ but with the opposite polarities. Because of charge segregation, small particles ($St_{\mathrm{S}}=1$) are negatively charged while large particles ($St_{\mathrm{L}}=10$) are positive. In previous measurements of tribocharged particles, the surface charge density could reach $\sigma_{\mathrm{p}} \approx 10^{-5} \ \mathrm{\mu C/m^2}$ \citep{WaitukaitisPRL2014, LeeNatPhys2015, HarperEPSL2021}. For particle size of $d_{\mathrm{p}} \sim 10 \ \mathrm{\mu m}$, the order of particle charge can be estimated as $q = \pi d_{\mathrm{p}}^2 \sigma_{\mathrm{p}} \sim 10^{-14} \ \mathrm{C}$. Moreover, since the considered particle concentration is low, particle collision is negligible. So the charge transfer is not included and the prescribed charge on each individual particle remains constant. 

\begin{table}
  \begin{center}
\def~{\hphantom{0}}
  \begin{tabular}{lll}
    Parameters & Values & Units \\
    \\
    \textit{Flow properties} \\
    Domain size, $L$ & $2 \pi \times 10^{-2}$ & $\mathrm{m}$ \\
    Grid number, $N_{\mathrm{grid}}^3$ & $192^3$ & $-$ \\
    Taylor Renolds number, $\mathrm{Re}_{\lambda}$ & $147.5$ & $-$ \\
    Fluid density, $\rho_\mathrm{f}$ & $1.0$ & $\mathrm{kg/m^3}$ \\
    Fluid kinetic viscosity, $\nu_\mathrm{f}$ & $1 \times 10^{-5}$ & $\mathrm{m^2/s}$ \\
    Fluctuation velocity, $u^\prime$ & $0.587$ & $\mathrm{m/s}$ \\
    Dissipation rate, $\epsilon$ & $8.18$ & $\mathrm{m^2/s^3}$ \\
    Kolmogorov length scale, $\eta$ & $1.05 \times 10^{-4}$ & $\mathrm{m}$ \\
    Kolmogorov time scale, $\tau_{\eta}$ & $1.11 \times 10^{-3}$ & $\mathrm{s}$ \\
    Integral length scale, $L_{\mathrm{int}}$ & $9.08 \times 10^{-3}$ & $\mathrm{m}$ \\
    Eddy turnover time, $\tau_{\mathrm{e}}$ & $ 1.547 \times 10^{-2}$ & $\mathrm{s} $\\
    \\
    \textit{Particle properties} \\
    Particle density, $\rho_{\mathrm{p}}$ & $1984.5$ & $\mathrm{kg/m^3}$ \\
    Particle diameter, $d_{\mathrm{p,S}} / d_{\mathrm{p,L}}$ & $10 / 31.6$ & $\mathrm{\mu m}$ \\
    Particle charge, $q$ & $\{0,0.5,1,2\} \times 10^{-14}$ & $\mathrm{C}$ \\
    Particle number, $N_{\mathrm{p,S}} / N_{\mathrm{p,L}}$ & $10^5 / 10^5$ & $-$\\
    Particle volume fraction, $\phi_{\mathrm{par}}$ & $6.87 \times 10^{-6}$ & $-$ \\

  \end{tabular}
  \caption{Simulation parameters.}
  \label{tab:SimulationParameters}
  \end{center}
\end{table}

\subsection{Fluid phase}
\label{sec:DNS}
The direct numerical simulation of the incompressible homogeneous isotropic turbulence is performed using the pseudo-spectral method. The computation domain is a triply periodic cubic box with the domain size $L^3={( 2 \pi \times 10^{-2} \ \mathrm{m})}^3$ and the grid number $N_{\mathrm{grid}}^3=192^3$. The governing equations of the fluid phase are given by

\begin{equation}
    \bnabla \bcdot \boldsymbol{u} = 0,
\end{equation}

\noindent
and

\begin{equation}
    \frac{\partial \boldsymbol{u}}{\partial t} + \boldsymbol{u} \bcdot \bnabla \boldsymbol{u} = -\frac{1}{\rho_{\mathrm{f}}}\bnabla p + \nu_{\mathrm{f}} \nabla^2 \boldsymbol{u} + \boldsymbol{f}_{{\mathrm{F}}}.
\end{equation}

\noindent
Here, $\boldsymbol{u}(\boldsymbol{x}_i)$ is the fluid velocity at the grid node $\boldsymbol{x}_i$, $p$ is the pressure, $\rho_{\mathrm{f}}$ is the fluid density, and $\nu_{\mathrm{f}}$ is the fluid kinematic viscosity. $\boldsymbol{f}_{\mathrm{F}}$ is the forcing term that injects kinetic energy from large scales and keeps the turbulence steady. In the wavenumber space, the forcing term is given by

\begin{equation}
    \hat{\boldsymbol{f}}_{\mathrm{F}}(\boldsymbol{\kappa}) = 
    \left\{
    \begin{aligned}
     & C \hat{\boldsymbol{u}}(\boldsymbol{\kappa}) &, |\boldsymbol{\kappa}| \le {\kappa}_\mathrm{crit} \\\
     & 0 &, |\boldsymbol{\kappa}| > {\kappa}_\mathrm{crit}
    \end{aligned}
    \right.
\end{equation}

\noindent
Here $C$ is the constant forcing coefficient, the critical wavenumber is $\kappa_\mathrm{crit} = 5\kappa_{0}$ with ${\kappa}_0 = 2 \pi / L$ the smallest wavenumber. 

When evolving the turbulence, the second-order Adam-Bashforth scheme is used for the nonlinear term while the viscous term is integrated exactly \citep{VincentJFM1991}. The fluid time step is $\Delta t_{\mathrm{F}}=1 \times 10^{-5} \ \mathrm{s}$. Table \ref{tab:SimulationParameters} lists typical parameters of the homogeneous and isotropic turbulence. The fluctuation velocity $u^\prime$ and the dissipation rate $\epsilon$ can be obtained from the energy spectrum $E(\kappa)$ as

\begin{equation}
    \frac{3}{2} {u^\prime}^2 = \int_{\kappa_{\mathrm{min}}}^{\kappa_{\mathrm{max}}}  E(\kappa) \mathrm{d} \kappa, \
    \epsilon = \int_{\kappa_{\mathrm{min}}}^{\kappa_{\mathrm{max}}} 2 \nu {\kappa}^2 E(\kappa) \mathrm{d} \kappa.
\end{equation}

\noindent
The Kolmogorov length and time scales are given by $\eta = ({\nu_{\mathrm{f}}}^3 / \epsilon)^{1/4}$ and $\tau_{\eta} = (\nu_{\mathrm{f}} / \epsilon)^{1/2}$, respectively. The integral length scale is $l= ({\pi}/2{u'}^2) \int_{\kappa_{\mathrm{min}}}^{\kappa_{\mathrm{max}}} [E(\kappa)/{\kappa}] \mathrm{d}{\kappa}$, the eddy turnover time is $\tau_{\mathrm{e}} = l/u^{\prime}$, and the Taylor Reynolds number is $\mathrm{Re}_{\lambda} = u^{\prime} \lambda / \nu_{\mathrm{f}}$ with $\lambda = u^{\prime} (15 \nu_{\mathrm{f}} /\epsilon)^{1/2}$ the Taylor microscale.

\subsection{Particle motion}
\label{sec:Partciel_Motion}
The movements of charged particles are integrated using the second-order Adam-Bashforth time stepping. The particle time step is $\Delta t_{\mathrm{p}} = 2.5 \times 10^{-6} \ \mathrm{s}$. This time step is much smaller than the particle relaxation time to well resolve particles' aerodynamic response to the turbulence ($\Delta t_{\mathrm{p}}/\tau_{\mathrm{p,S}}=0.23 \%$). Besides, within each step, particles only move a short distance compared to the Kolmogorov length ($u^{\prime} \Delta t_{\mathrm{p}} / \eta = 1.4 \%$), so that the variation of the electrostatic force is well captured when particles are getting close. The governing equations of particle motions are 

\begin{subequations}
\begin{equation}
\label{eq:ParticleMotion}
m_i \frac{\mathrm{d} \boldsymbol{v}_i}{\mathrm{d} t} = \boldsymbol{F}^{\mathrm{F}}_i + \boldsymbol{F}^{\mathrm{E}}_{i},
\end{equation}

\begin{equation}
\frac{\mathrm{d} \boldsymbol{x}_i}{d t} = \boldsymbol{v}_i.
\end{equation}
\end{subequations}

\noindent
Here, $\boldsymbol{v}_i$ and $\boldsymbol{x}_i$ are the velocity and the location of particle $i$, respectively. $m_i = \pi \rho_{\mathrm{p}} d_{\mathrm{p},i}^3/6$ is the particle mass, with $\rho_{\mathrm{p}}$ the density and $d_{\mathrm{p},i}$ the diameter. $\boldsymbol{F}^{\mathrm{F}}_i$ and $\boldsymbol{F}^{\mathrm{E}}_i$ are the fluid force and the electrostatic force acting on particle $i$.

Since particles are small ($d_{\mathrm{p}} < \eta$) and heavy ($\rho_{\mathrm{p}} / \rho_{\mathrm{f}} \gg 1$), the dominant fluid force comes from the Stokes drag \citep{MaxeyPOF1983} 

\begin{equation}
    \boldsymbol{F}^{\mathrm{F}}_i = -3 \pi \mu_{\mathrm{f}} d_{\mathrm{p}} (\boldsymbol{v}_i - \boldsymbol{u}( \boldsymbol{x}_{i})).
\end{equation}

\noindent
Here $\mu_{\mathrm{f}}$ is the dynamic viscosity of the fluid, and $\boldsymbol{u}(\boldsymbol{x}_{i})$ is the fluid velocity at the particle position. For two consecutive particle time steps within the same fluid time step, the flow field remains unchanged, and the fluid velocity at the particle location is interpolated at each particle time step using the second-order tri-linear interpolation scheme \citep{MarshallJCP2009, QianPRE2022}. The effect of fluid inertia is neglected by assuming that the particle Reynolds number $\mathrm{Re}_{\mathrm{p}}$ is much smaller than unity.

In this dilute system, the average separation between charged particles is large compared to particle size ($\Bar{l} \gg d_{\mathrm{p}}$) indicating insignificant effect of particle polarization \citep{RuanJCP2022}. Therefore, the electrostatic force acting on each particle $i$ can be obtained by summing the pairwise Coulomb force from other particles $j$, which reads

\begin{equation}
\label{eq:CoulombForce}
\boldsymbol{F}_{i}^{\mathrm{E}} = \sum_{j \ne i} \frac{q_i q_j 
 (\boldsymbol{x}_i - \boldsymbol{x}_j)}{4 \pi \varepsilon_0 {|\boldsymbol{x}_i - \boldsymbol{x}_j|}^3}.
\end{equation}

\noindent
Here, $\varepsilon_0 = 8.8542 \times 10^{-12} \ \mathrm{F/m}$ is the vacuum permittivity, $q_i$ is the point charge located at the centroid of particle $i$. 

To properly apply the periodic boundary condition of the long-range electrostatic force, several layers of image domains are added around the original computation domain \citep{DiRenzo2018, BoutsikakisJCP2023}. When calculating the electrostatic force acting on the $i$th particle located at $\boldsymbol{x}_i$ , the contributions from other particles $j$ at $\boldsymbol{x}_j$ as well as their images at $\boldsymbol{x}_j +(l\boldsymbol{i}+m\boldsymbol{j}+n\boldsymbol{k})L$ are all considered:

\begin{equation}
\boldsymbol{F}_{i}^{\mathrm{E}} = \sum_{j \ne i} \sum_{l,m,n} \frac{q_i q_j  \{ \boldsymbol{x}_i - [\boldsymbol{x}_j+(l\boldsymbol{i}+m\boldsymbol{j}+n\boldsymbol{k})L ] \} }{4 \pi \varepsilon_0 { \left| \boldsymbol{x}_i - [\boldsymbol{x}_j+(l\boldsymbol{i}+m\boldsymbol{j}+n\boldsymbol{k})L] \right|}^3},
\label{eq:PeridoicCoulombForce}
\end{equation}

\noindent
where $l,m,n=-N_{\mathrm{per}}, ..., N_{\mathrm{per}}$. $\boldsymbol{i}$, $\boldsymbol{j}$, $\boldsymbol{k}$ are unit vectors along the $x$, $y$, and $z$ directions.

According to Eq. \ref{eq:PeridoicCoulombForce}, when computing the electrostatic force acting on the target particle $i$, the cost is $O(N_{\mathrm{s}}) = O((N_{\mathrm{p}}-1)\times(2N_{\mathrm{per}}+1)^3)$, where $O(N_{\mathrm{s}})$ is the number of source particles to each target particle. The total cost of electrostatic computation then scales with $O(N_{\mathrm{p}}^2 N_{\mathrm{per}}^3)$. Considering the large particle number and the addition of image domains, direct summating Eq. \ref{eq:PeridoicCoulombForce} is extremely expensive. Therefore, we employ the fast multipole method (FMM) to accelerate the calculation. In FMM, the field strength generated by neighboring particles are directly computed using Eq. \ref{eq:PeridoicCoulombForce}, while the contribution from clouds of particles far from the target particle is approximated using fast multipole expansion, which reduces the required computation cost to $O(N_{\mathrm{p}}\mathrm{log}(N_{\mathrm{p}}N_{\mathrm{per}}^3))$ \citep{GreengardJCP1987, GreengardActa1997}. In this study the open-source library FMMLIB3D is employed to conduct fast electrostatic computation \citep{GimbutasCCP2015}. The accuracy of FMM with various layers of image domains is compared in Appendix \ref{appA}. Based on the analysis, two layers of image boxes ($N_{\mathrm{per}}=2$) are added, which is sufficient to ensure the convergence of the electrostatic force acting on all particles in the original domain. 

Moreover, the Coulomb force is capped if the distance between two particles is smaller than a preset critical distance to avoid nonphysically strong force:

\begin{equation}
F_{i \leftarrow j}^{\mathrm{E}} = \mathrm{min} \left \{ \frac{q_i q_j}{4 \pi \varepsilon_0 {|\boldsymbol{x}_i - \boldsymbol{x}_j|}^2}, \frac{q_i q_j}{4 \pi \varepsilon_0 {d_{\mathrm{cap}}}^2} \right \}.
\end{equation}

\noindent
Here the critical distance is $d_{\mathrm{cap}} = d_{\mathrm{p,SS}}/2$ or $d_{\mathrm{cap}} = \eta /20$. This value is chosen so that even for a pair of small particles approaching each other, the Coulomb force between them is still accurate at the collision distance ($|\boldsymbol{x}_i - \boldsymbol{x}_j| = d_{\mathrm{p,SS}}>d_{\mathrm{cap}}$). Moreover, since $d_{\mathrm{cap}}$ is set small compared to the Kolmogorov length scale $\eta$, capping the electrostatic force does not affect the statistics at larger scales presented below.

In each run, the single-phase turbulence is first evolved until reaching the steady state. Particles are then injected randomly in the domain with their initial velocities equal to the fluid velocity at particle locations. It takes roughly $3.5 \tau_{\mathrm{e}}$ for particle dynamics to reach equilibrium, statistics are then taken over another $5\tau_{\mathrm{e}}$. Three parallel runs with different initial particle locations are performed for each case, and their results are averaged and presented below.

\section{Results and discussions}
\label{sec:Results_and_Discussions}

\subsection{Clustering and relative velocity of charged particles}
\label{sec:1}

\begin{figure}
    \centering
    \includegraphics[width=12cm]{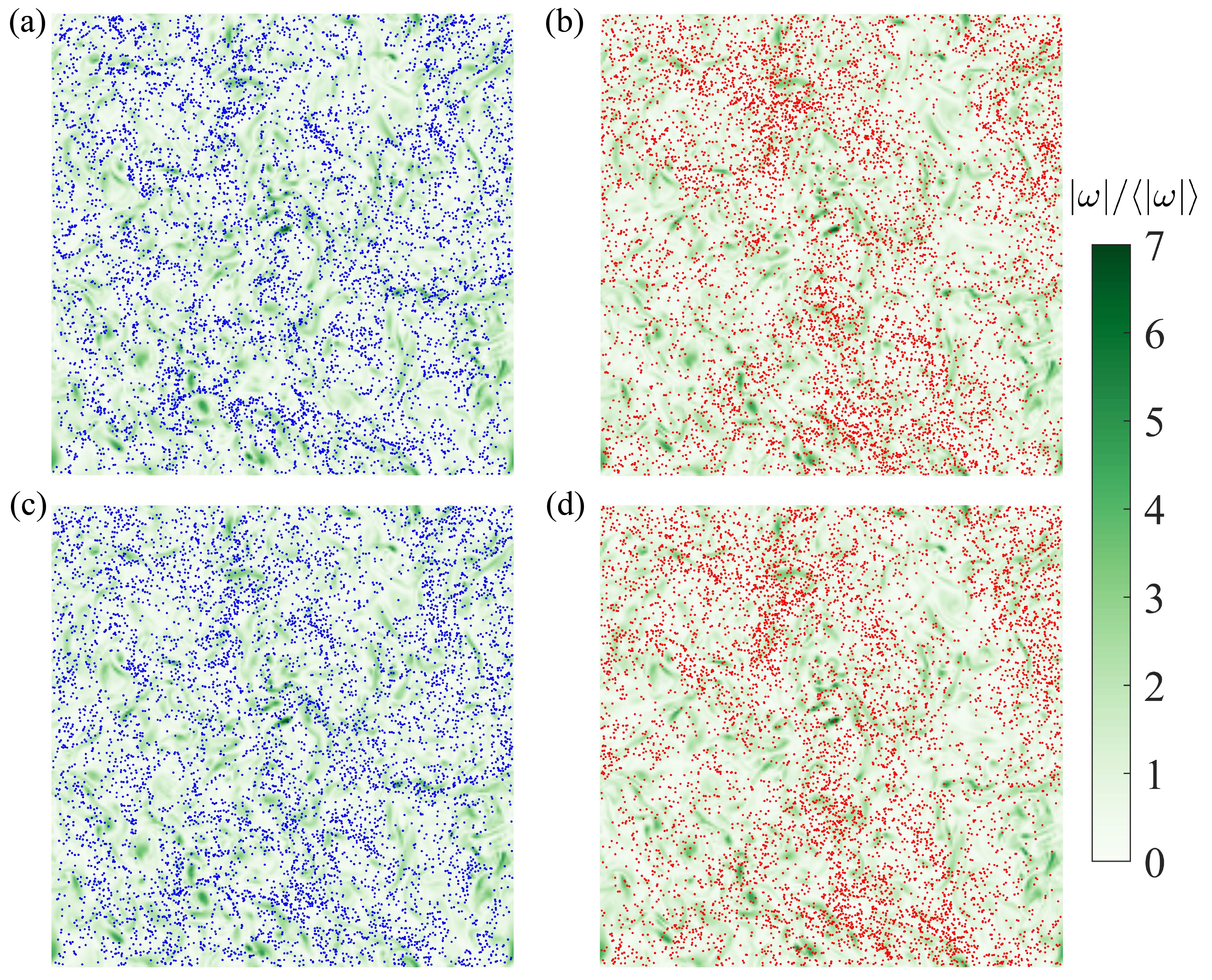}
    \caption{Snapshots of particles in the slice $|z| \le 20 \eta$ with (a) $St=1$ and $q = 0 \ \mathrm{C}$, (b) $St=10$ and $q = 0 \ \mathrm{C}$, (c) $St=1$ and $q = 2 \times 10^{-14} \ \mathrm{C}$, (d) $St=10$ and $q = 2 \times 10^{-14} \ \mathrm{C}$. Blue/red dots represent small/large particles. The color bar indicates the vorticity magnitude normalized by its average over the domain.}
    \label{fig:1}
\end{figure}

Figure \ref{fig:1} compares the spatial distribution of bidispersed particles within a thin slice $|z| \leq 20 \eta$. Although oppositely-charged bidispersed particles are simulated in each case, we show small and large particles in the left and the right panels separately for better illustration. In the neutral case (Fig.\ref{fig:1}(a) and (b)), particle behavior is solely determined by the particle-turbulence interaction. Small particles ($St=1$) are responsive to fluctuations at the Kolmogorov length scale. As a result, their spatial distribution is highly nonuniform, and small-scale particle clusters can be observed (Fig.\ref{fig:1}(a)). Meanwhile, large particles ($St=10$) are more inertial, so they are more dispersed in the domain (Fig.\ref{fig:1}(b)). 

We employ the radial distribution function (RDF) to quantify how particles from the same (or different) size groups cluster. The radial distribution functions of different kinds of particle pairs, i.e., small-small (SS), large-large (LL), and small-large (SL) paris, are defined as 

\begin{subequations}
    \begin{equation}
        g_{\mathrm{SS}}(r) = \frac{\Delta N_{\mathrm{SS}}(r) / \Delta V(r)}{N_{\mathrm{p,S}}(N_{\mathrm{p,S}}-1)/2/L^3}, \ 
        g_{\mathrm{LL}}(r) = \frac{\Delta N_{\mathrm{LL}}(r) / \Delta V(r)}{N_{\mathrm{p,L}}(N_{\mathrm{p,L}}-1)/2/L^3},
    \end{equation}

    \begin{equation}
        g_{\mathrm{SL}}(r) = \frac{\Delta N_{\mathrm{SL}}(r) / \Delta V(r)}{N_{\mathrm{p,S}}N_{\mathrm{p,L}}/L^3}.
    \end{equation}
\end{subequations}

\noindent
In the definition of $g_{\mathrm{SS}}(r)$, $\Delta N_{\mathrm{SS}}(r)$ is the number of SS pairs with their separation distances lie within the range of $r \pm \Delta r/2$, and $\Delta V(r) = 4 \pi [(r+\Delta r/2)^3-(r-\Delta r/2)^3]/3$ is the shell volume in the separation distance bin. $N_{\mathrm{p,S}}(N_{\mathrm{p,S}}-1)/2/L^3$ is the average number of SS pairs in the whole domain. Therefore, $g_{\mathrm{SS}}(r)>1$ means there is a higher density of SS pairs at a given separation distance of $r$ compared to the average pair density. The definitions of $g_{\mathrm{LL}}(r)$ and $g_{\mathrm{SL}}(r)$ are similar and thus omitted.

\begin{figure}
    \centering
    \includegraphics[width=13.5cm]{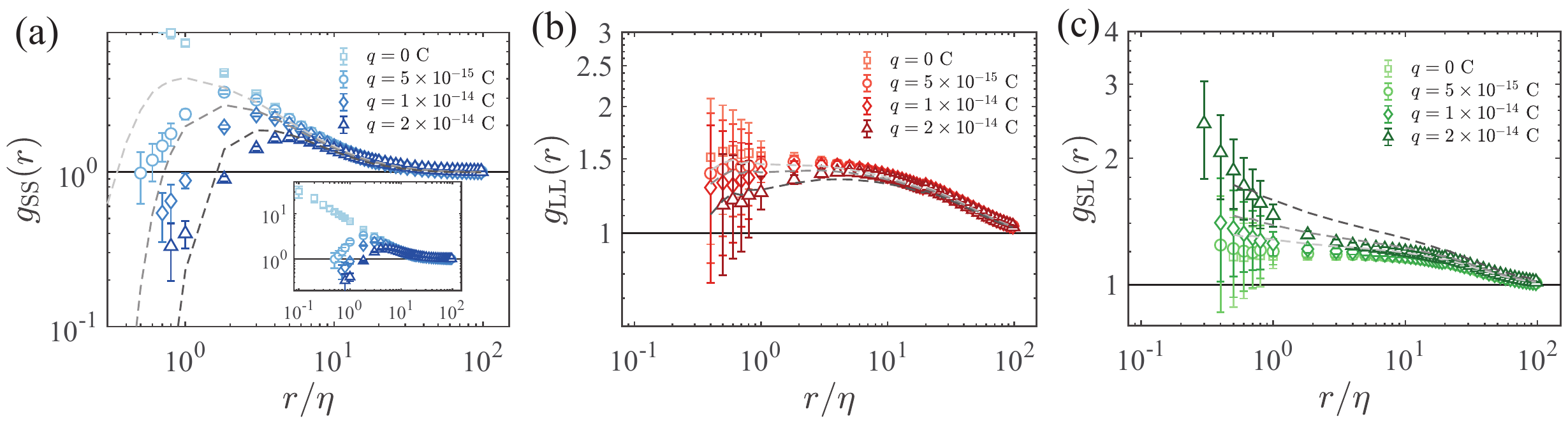}
    \caption{RDFs between (a) small-small, (b) large-large, and (c) small-large particle pairs with different particle charge $q$. The inset in (a) shows the full scale of $g_\mathrm{SS}(r)$ in the neutral case. Dashed lines from light to dark correspond to predicted RDFs using Eq. \ref{eq:gr_model_SS_final} (or its counterpart for the oppositely-charged case) for $q=5 \times 10^{-15} \ \mathrm{C}$, $1 \times 10^{-14} \ \mathrm{C}$, and $2 \times 10^{-14} \ \mathrm{C}$, respectively.}
    \label{fig:2}
\end{figure}

As shown in Fig. \ref{fig:2}(a), $g_{\mathrm{SS}}(r)$ for neutral pairs is significantly larger than unity at $r \leq \eta$, indicating strong clustering at the small scale. $g_{\mathrm{SS}}(r)$ also follows a clear power law $g_{\mathrm{SS}}(r) \sim (r/ \eta)^{-\mathrm{c}_1}$ with the fitted exponent $\mathrm{c}_1 = 0.69$. The value of  $\mathrm{c}_1$ is close to those reported in previous studies  with similar $\mathrm{Re}_{\lambda}$, e.g., $\mathrm{c}_1 = 0.67$ for $\mathrm{Re}_{\lambda}=143$ \citep{SawNJP2012} and $\mathrm{c}_1 = 0.68$ for $\mathrm{Re}_{\lambda}=140$  \citep{IrelandJFM2016a}. In comparison, $g_{\mathrm{LL}}(r)$ for neutral pairs is only slightly larger than one, which means the clustering of large particles are relatively weak. However, $g_{\mathrm{LL}}(r)$ decreases slowly with the increase of $r$ and stays above unity until $r \approx 100 \eta$, while  $g_{\mathrm{SS}}(r)$ drops rapidly and approaches unity around $20 - 30 \eta$. This is because the clustering of inertial particles with the relaxation time $\tau_{\mathrm{p}}$ are driven by vortices whose timescales are comparable to particle's relaxation time (i.e., $\tau(r) \sim \tau_{\mathrm{p}}$). In the inertial range, the time scale satisfy $\tau(r)\sim \epsilon^{-1/3} r^{2/3}$, so the relationship can be given as $r \sim \epsilon^{1/3} \tau_{\mathrm{p}}^{3/2}$ \citep{YoshimotoJFM2007, BecJPCS2011}. With the increase of particle inertia, the length and the time scales of vortices that could affect particle behaviors also become larger. As a result, the size of particle clusters also grow larger, leading to the large correlation length in $g_{\mathrm{LL}}(r)$ \citep{YoshimotoJFM2007, LiuJFM2020}. Besides, when comparing those particles with large St difference, because their relaxation times differ by a factor of ten, they respond to flows of very different scales. Therefore, there is little spatial correlation between small and large particles, and $g_{\mathrm{SL}}(r)$ is close to unity at all separation distance $r$ when particles are neutral.

When particles are charged, since charge segregation correlates with size, the same-size particles will repel each other when they get close. Since the amount of particle charge is the same, the effect of repulsion is more drastic between SS pairs rather than between LL pairs. As a result, the order of $g_{\mathrm{SS}}(r)$ at small $r$ drops by several decades as $q$ increases. The same decreasing trend is observed for $g_{\mathrm{LL}}(r)$, but the influence is less significant because of the large particle inertia. This can also be shown by comparing the top and the bottom panels in Fig. \ref{fig:1}. In the charged case ($q=2 \times 10^{-14} \ \mathrm{C}$), the small particles become less concentrated (Fig. \ref{fig:1}(c)), while no obvious difference is seen in the spatial distribution of large particles (Fig. \ref{fig:1}(d)). It is worth noting that, since the Coulomb force decays rapidly with $r$, its effect is only significant within a relatively short range ($r \approx 1-10 \eta$). Beyond this range, both $g_{\mathrm{SS}}(r)$ and $g_{\mathrm{LL}}(r)$ in Fig.\ref{fig:2} (a) and (b) recover to their neutral values again, so clustering could still be observed at large $r$. As for $g_{\mathrm{SL}}$, because of the Coulomb attraction, particles of different sizes are more likely to stay close. More specifically, considering the large mass difference between SL pairs ($m_{\mathrm{L}}/m_{\mathrm{S}}=(St_{\mathrm{L}}/St_{\mathrm{S}})^{1.5}\approx30$), small particles will always get attracted towards the large ones, giving rise to the rapid growth of $g_{\mathrm{SL}}(r)$ at small separation ($r \leq \eta$). Again, the opposite-sign attraction decays with increasing $r$, so $g_{\mathrm{SL}}(r)$ approaches one when $r$ is sufficiently large.

Apart from spatial correlation, it is also of interest to study the relative velocity between particle pairs and focus on the modulation caused by the electrostatic force. For a pair of particles $i$ and $j$ with the separation $r = |\boldsymbol{x}_i-\boldsymbol{x}_j|$, the radial component of the relative velocity is defined as $w_{\mathrm{r},ij} = (\boldsymbol{v}_i-\boldsymbol{v}_j) \cdot (\boldsymbol{x}_i-\boldsymbol{x}_j)/|\boldsymbol{x}_i-\boldsymbol{x}_j|$. Taking the ensemble average over SS/LL/SL particle pairs then yields the mean relative velocities between three kinds of particle pairs as $\langle |w_{\mathrm{r,SS}}| \rangle$, $\langle |w_{\mathrm{r,LL}}| \rangle$, and $\langle |w_{\mathrm{r,SL}}| \rangle$, respectively.

\begin{figure}
    \centering
    \includegraphics[width=13.5cm]{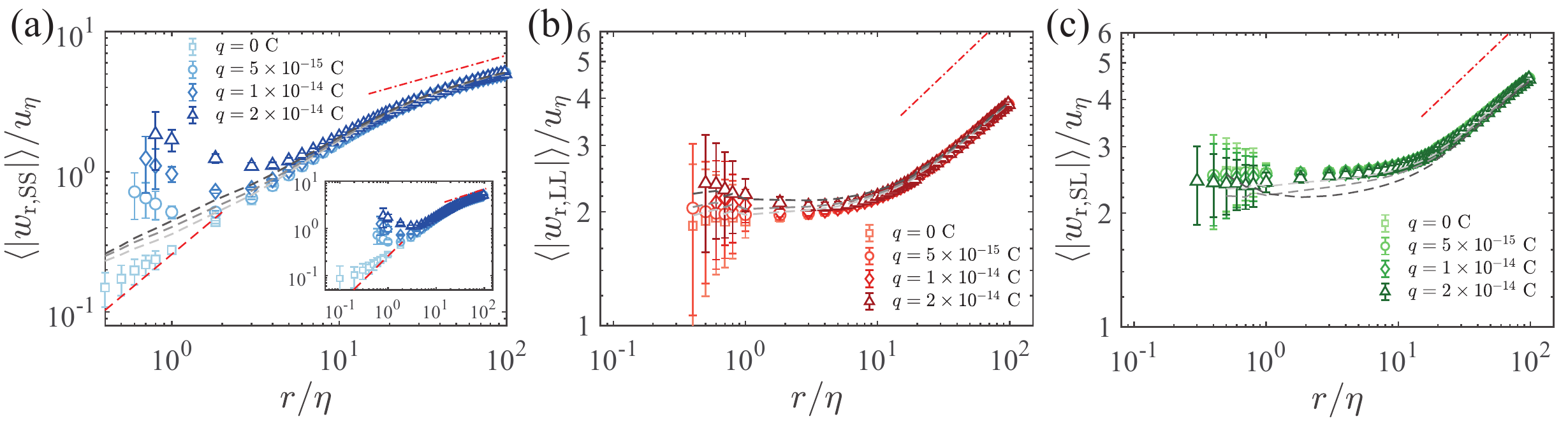}
    \caption{Mean radial relative velocity $ \langle |w_{\mathrm{r}}| \rangle$ normalized by $u_{\eta}$ between (a) small-small, (b) large-large, and (c) small-large particle pairs with different particle charge $q$. The inset in (a) shows the full scale of$ \langle |w_{\mathrm{r,SS}}| \rangle$ in the neutral case. The red dashed line denotes the velocity scaling $ \propto r$ in the dissipation range, and the red dash-dotted line denotes the velocity scaling $\propto r^{1/3}$ in the inertial range. Dashed lines from light to dark correspond to predictions using Eq. \ref{eq:wr_model_SS} (or its counterpart for the oppositely-charged case) for $q=5 \times 10^{-15} \ \mathrm{C}$, $1 \times 10^{-14} \ \mathrm{C}$, and $2 \times 10^{-14} \ \mathrm{C}$, respectively. }
    \label{fig:3}
\end{figure}

Fig. \ref{fig:3} (a) compares the mean radial relative velocity between SS pairs, $\langle |w_{\mathrm{r,SS}}| \rangle$, for various particle charge $q$. For neutral SS pairs, when $r$ is in the inertial range ($\eta \ll r \ll L_{\mathrm{int}}$), $\tau_{\mathrm{p,S}}$ is much smaller than the characteristic time scales of turbulent fluctuations at this length scale. The radial relative velocity between SS pairs, $\langle |w_{\mathrm{r,SS}}| \rangle$, thus follows that of fluid tracers $\delta u_{\mathrm{r}}$. Here, the fluid relative velocity is obtained from the the second order longitudinal structure function as $\delta u_{\mathrm{r}} = {C_2}^{1/2} \epsilon^{1/3} r^{1/3}$ and the constant is $C_2=2.13$ \citep{YeungPRE1997} (shown as dash-dotted lines in Fig. \ref{fig:3}). If $r$ is within the dissipative range, the timescale of local fluctuations becomes comparable to $\tau_{\mathrm{p,S}}$. Small neutral particles cannot perfectly follow the background flow at this separation, so $\langle |w_{\mathrm{r,SS}}| \rangle$ deviates from the fluid relative velocity $\delta u_{\mathrm{r}}= (\epsilon/15 \nu)^{1/2} r$ (dashed lines in Fig.\ref{fig:3}). 

When particles are charged, since the influence of Coulomb force is negligible at large $r$, curves for different $q$ collapse. In contrast, $\langle |w_{\mathrm{r,SS}}| \rangle$ is found to rise significantly as particles get close to each other. When $q$ gets larger, the increase of $\langle |w_{\mathrm{r,SS}}| \rangle$ becomes more pronounced and the deviation from the neutral result also occurs at a larger $r$. This seems counter-intuitive because the Coulomb repulsion between SS pairs should always slow approaching particles down and reduce the relative velocity. One explanation for this change is as follows. When $q$ is large, the strong Coulomb repulsion causes biased sampling at small $r$: only those pairs with large inward relative velocity could overcome the energy barrier and get close. Meanwhile, since the electrical potential energy is conserved in the approach-then-depart process, the pairs approaching each other at a high velocity will also be repelled away at a high velocity as the electrostatic force pushing them apart. As a result, pairs with a large $|w_{\mathrm{r,SS}}|$ take a large proportion as $q$ increase, so the mean radial relative velocity between both approaching and departing pairs, $\langle w_{\mathrm{r},ij} \rangle$, shows the increasing trend in Fig. 3(a). This increasing trend with $q$ will be further discussed in Sec \ref{sec:3}.

Compared to small particles with $St=1$, large particles with $St=10$ are very inertial and insensitive to fluctuations even at large separations. The curves of $\langle |w_{\mathrm{r,LL}}| \rangle$ are therefore much flatter. Besides, the constant relative velocity between LL pairs at small $r$ is also significantly larger, because their relative velocity comes from the energetic large-scale motions. When particles are charged, the same increasing trend with $q$ is also observed but less significant, which can be attributed to the large kinetic energy compared to the electrostatic potential energy. 

As for SL pairs, $\langle |w_{\mathrm{r,SL}}| \rangle$ is larger than both $\langle |w_{\mathrm{r,SS}}| \rangle$ and $\langle |w_{\mathrm{r,LL}}| \rangle$ in the neutral cases. This is caused by the different responses of small/large particles to turbulent fluctuations, which is termed as the differential inertia effect \citep{ZhouJFM2001}. Interestingly, there is no obvious change when comparing curves between different charge $q$ in Fig.\ref{fig:3}(c). As shown in Fig. \ref{fig:2}(c), Coulomb force attracts small particles towards large ones and enhances their spatial correlation. Nevertheless, even though charged SL pairs experience a more similar flow history than neutral SL pairs, they will still develop a large relative velocities over time because their response to local fluctuations is very different. Therefore, the influence of charge on $\langle |w_{\mathrm{r,SL}}| \rangle$ is weak.
 
\subsection{Effect of Coulomb force on particle flux}
\label{sec:2}

Once the radial distribution function and the radial relative velocity are known, we could further investigate the collision frequency between charged particles. For a steady system, the collision frequency can be measured by the kinematic collision kernel function as \citep{SundaramJFM1997, WangJFM2000}

\begin{equation}
    \Gamma_{ij}(R_{\mathrm{C}}) = 2 \pi R_{\mathrm{C}}^2 \cdot g_{ij}(R_{\mathrm{C}}) \cdot \langle |w_{\mathrm{r},ij}|\rangle (R_{\mathrm{C}}).
\end{equation}

\noindent
Here, $R_{\mathrm{C}}=r_{i}+r_{j}$ is the collision radius, which equals to the sum of the radii of particles $i$ and $j$. $g_{ij}(R_{\mathrm{C}})$ is the RDF at $r=R_{\mathrm{C}}$, and the mean relative velocity is $\langle |w_{\mathrm{r},ij}|\rangle = \int_{-\infty}^{\infty} |w_{\mathrm{r},ij}| p_{ij}(w_{\mathrm{r},ij}) \mathrm{d}w_{\mathrm{r},ij}$ with $p_{ij}(w_{\mathrm{r},ij})$ the probability density function (PDF) of $w_{\mathrm{r},ij}$ at $r=R_{\mathrm{C}}$.

Even though other particle parameters (e.g., the Stokes number) are set the same, different choices of $R_{\mathrm{C}}$ will affect the final outcome of $\Gamma_{ij}$ by changing the collision geometry. Therefore, instead of directly using $\Gamma_{ij}$, we define the particle flux $\Phi_{ij}$ as the ratio of the collision kernel $\Gamma_{ij}(r)$ to the area of the collision sphere $4 \pi r^2$ at $r$ as:

\begin{equation}
\label{eq:FullFlux}
    \Phi_{ij}(r) = \frac{\Gamma_{ij}(r)}{4 \pi r^2} = \frac{1}{2} g_{ij}(r) \cdot \langle |w_{\mathrm{r},ij}|\rangle (r).
\end{equation}

\noindent
$\Phi_{ij}$ can be understood as the number of particles crossing the collision sphere per area per unit time, which is independent of the prescribed $R_{\mathrm{C}}$.

Apart from Eq.\ref{eq:FullFlux} that uses information of both approaching and departing pairs, the particle flux can also be defined using only the approaching or departing pairs. In real situations, it is the approaching pairs that lead to collisions, so a natural way to define particle flux is based on the inward flux as

\begin{equation}
\label{eq:InwardFlux}
    \Phi_{ij}^{\mathrm{in}}(r) = g_{ij}(r) F(w_{\mathrm{r},ij}<0|r) \cdot \langle w_{\mathrm{r},ij}^{-}\rangle (r).
\end{equation}

\noindent
Here, $F(w_{\mathrm{r},ij}<0|r)$ is the fraction of particles moving inwards at $r$, and the mean inward relative velocity is $\langle w_{\mathrm{r},ij}^{-}\rangle = - \int_{-\infty}^{0} w_{\mathrm{r},ij} p_{ij}(w_{\mathrm{r},ij}) \mathrm{d}w_{\mathrm{r},ij}$. After the system reaches the equilibrium, RDFs between particle pairs no longer change, suggesting that the inward flux should balance the outward flux at all $r$. Here,
the outward particle flux is

\begin{equation}
\label{eq:OutwardFlux}
    \Phi_{ij}^{\mathrm{out}}(r) = g_{ij}(r) F(w_{\mathrm{r},ij}>0|r) \cdot \langle w_{\mathrm{r},ij}^{+}\rangle (r),
\end{equation}

\noindent
with the mean outward relative velocity given by $\langle w_{\mathrm{r},ij}^{+}\rangle = \int_{0}^{\infty} w_{\mathrm{r},ij} p_{ij}(w_{\mathrm{r},ij}) \mathrm{d}w_{\mathrm{r},ij}$. The definition of $\Phi_{ij}$ is based on a pair of particles $i$ and $j$ without specifying their sizes. If the average is taken over all SS pairs, the result is the small-small particle flux denoted by $\Phi_{\mathrm{SS}}$. The flux between LL pairs ($\Phi_{\mathrm{LL}}$) and SL pairs ($\Phi_{\mathrm{SL}}$) can also be obtained by taking the average over corresponding particle pairs. 

The SS fluxes $\Phi_{\mathrm{SS}}$ defined by Eqs. \ref{eq:FullFlux}, \ref{eq:InwardFlux} and \ref{eq:OutwardFlux} are first compared and show good agreement with each other, indicating that the flux-balance condition is valid. Since Eq.\ref{eq:FullFlux} uses the information of both approaching and  departing particle pairs, it is adopted in the following sections for better statistics.

\begin{figure}
    \centering
    \includegraphics[width=13.5cm]{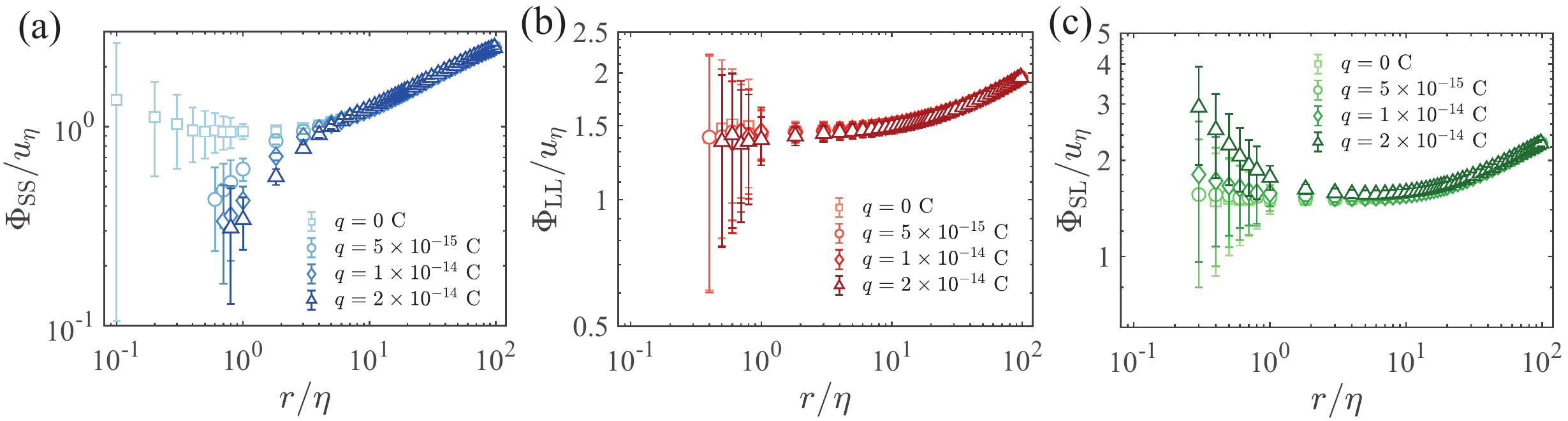}
    \caption{Particle fluxes between (a) small-small, (b) large-large, and (c) small -large particles pairs with different particle charge $q$.}
    \label{fig:4}
\end{figure}

Fig.\ref{fig:4}(a) compares the SS fluxes with various $q$. The flux for neutral pairs remain constant when $r$ is comparable to $\eta$. This indicates that, although both $g_{\mathrm{SS}}$ (Fig.\ref{fig:2}(a)) and $\langle |w_{\mathrm{r,SS}}| \rangle$ (Fig.\ref{fig:3}(a)) keep changing as $r$ decreases, their product is almost constant for small $r$. For inertial particle pairs with small $r$, it has been shown that $g(r) \propto r^{D_2-3}$ with $D_2$ the correlation dimension \citep{BecPRL2007}, while the relative velocity follows $\langle w_{\mathrm{r,SS}} \rangle \propto r^{3-D_2}$ \citep{GustavssonJOT2014}. Therefore, the dependence of $\Phi_{\mathrm{SS}}$ on $r$ is canceled out. Besides, in practical situations the collision diameter $R_{\mathrm{C}}$ between micron particles/droplets is smaller than $\eta$, so the constant flux at such separation distance leads to a quadratic dependence of collision kernel on $R_{\mathrm{C}}$ as $\Gamma_{SS}=4 \pi R_{\mathrm{C,SS}}^2 \Phi_{\mathrm{SS}}(R_{\mathrm{C,SS}})$ \citep{SundaramJFM1997}. 

When particles are charged, the SS flux is found to decrease rapidly as $r$ drops. To characterize the effect of Coulomb force on the reduction of $\Phi_{\mathrm{SS}}$, we need to quantify the competition between the driving force and the resistance of particle collisions. For a pair of same-sign particles $i$ and $j$ separating by $r$, the driving force can be evaluated by the relative kinetic energy $E_{\mathrm{Kin}}(r)=m_{ij} \langle |w_{\mathrm{r},ij}| \rangle^2/2$. Here, $m_{ij}=(m_i^{-1}+m_j^{-1})^{-1}$ is the effective mass, $\langle |w_{\mathrm{r},ij}| \rangle (r)$ is the mean radial relative velocity between neutral pairs with a separation of $r$. The resistance is the electrical energy barrier $E_{\mathrm{Coul}} = q_i q_j /4 \pi \varepsilon_0 r$. The energy ratio is then given as

\begin{equation}
\label{eq:Ct_0}
    \mathrm{Ct}_0 \equiv \frac{E_{\mathrm{Coul}}}{E_{\mathrm{Kin}}} = \frac{q_i q_j}{2 \pi \varepsilon_0 r m_{ij} \langle |w_{\mathrm{r},ij}| \rangle^2}.
\end{equation}

\noindent
Note that in previous studies regarding the clustering of charged particles with weak inertia \citep{LuPRL2010,LuNJP2010}, the approaching velocity between particle pairs can be directly derived through perturbation expansion in Stokes number \citep{ChunJFM2005}. In this work, however, particles are very inertial, so the mean relative velocity $\langle |w_{\mathrm{r},ij}| \rangle$ between neutral pairs is used instead. By using the corresponding effective mass (e.g., $m_{\mathrm{SS}}=m_{\mathrm{S}}/2$, $m_{\mathrm{LL}}=m_{\mathrm{L}}/2$, and $m_{\mathrm{SL}}=m_{\mathrm{S}}m_{\mathrm{L}}/(m_{\mathrm{S}}+m_{\mathrm{L}})$) and the mean relative velocity ($\langle|w_{\mathrm{r,SS}}|\rangle$, $\langle|w_{\mathrm{r,LL}}|\rangle$, and $\langle|w_{\mathrm{r,SL}}|\rangle$), Eq. \ref{eq:Ct_0} can quantify the Coulomb-turbulence competition for different kinds of particle pairs. Since $\mathrm{Ct}_0$ measures the relative importance of the Coulomb force to the mean relative kinetic energy, it is called the \emph{mean} Coulomb-turbulence parameter hereinafter.

\begin{figure}
    \centering
    \includegraphics[width=13.5cm]{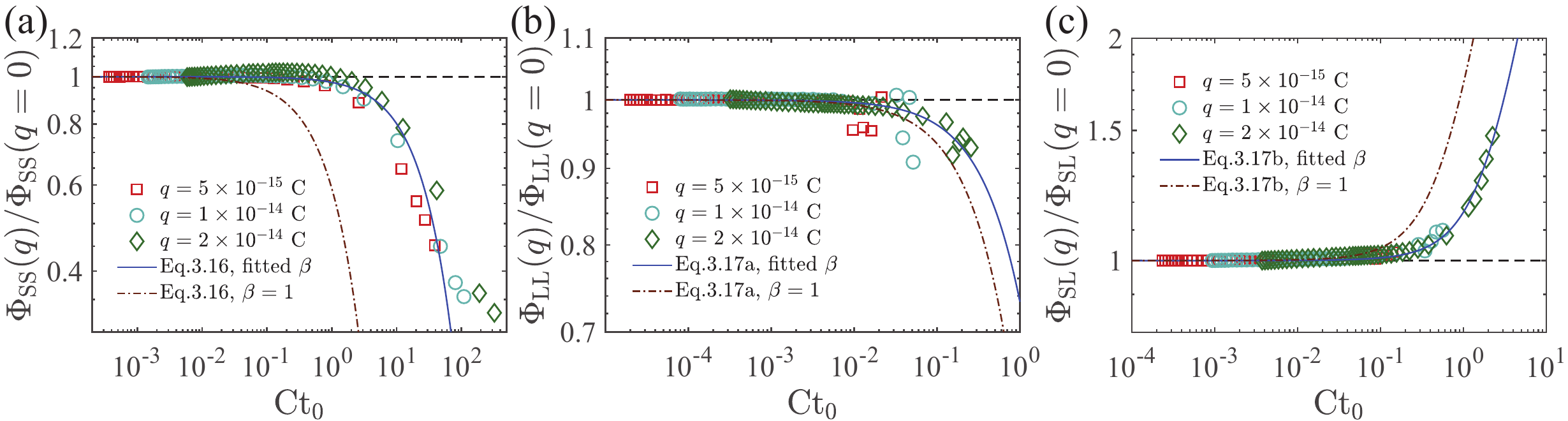}
    \caption{Ratios of charged particle flux $\Phi(q)$ to neutral particle flux $\Phi(q=0)$ as a function of the mean Coulomb-turbulence parameter $\mathrm{Ct}_0$ for (a) small-small, (b) large-large, and (c) small-large particle pairs. Blues solid lines are model predictions from Eqs. \ref{eq:FluxRatioModel_SS} and \ref{eq:FluxRatioModel_LL_SL} with fitted $\beta$. Brown dash-dotted lines are predictions with the fixed $\beta=1$. }
    \label{fig:5}
\end{figure}

According to the definition in Eq.\ref{eq:Ct_0}, the value of $\mathrm{Ct}_0$ can be varied as the particle charge or the separation distance changes. Therefore, for each data point in Fig.\ref{fig:4} (a), we plot the ratio of $\Phi_{\mathrm{SS}}(q)$ to its corresponding value in the neutral case $\Phi_{\mathrm{SS}}(q=0)$ as a function of $\mathrm{Ct}_0$ in Fig.\ref{fig:5}(a). The flux ratios for different $q$ are found to follow the same trend. The particle flux of charged particles is almost unaffected when $\mathrm{Ct}_0$ is small, and a significant decay is observed when $\mathrm{Ct}_0$ exceeds unity. The LL fluxes $\Phi_{\mathrm{LL}}$ in Fig.\ref{fig:4}(b) are analyzed in a similar way, and the results are plotted in Fig.\ref{fig:5}(b). Because of the large kinematic energy between LL pairs, the Coulomb repulsion is relatively weak ($\mathrm{Ct}_0<1$), so the reduction of $\Phi_{\mathrm{LL}}$ has not yet reached the electrostatic-dominant regime. As for the flux between SL pairs shown in Fig.\ref{fig:4}(c), the opposite trend is seen when plotting the flux ratio as a function of $\mathrm{Ct}_0$ (Fig.\ref{fig:5}(c)). The increase of $\Phi_{\mathrm{SL}}$ becomes significant when $\mathrm{Ct}_0$ becomes larger than one.

We now propose a model to quantify the influence of the electrostatic force on the mean particle flux. For particle pairs with a separation distance $r$, the mean radial relative velocities between different kinds of particle pairs have been shown in Fig.\ref{fig:3}. We then assume that, for each kind of particle pairs, the relative velocity between neutral pairs follows the Gaussian distribution. Taking SS particle pairs as an example, the probability density function $p_{\mathrm{r,SS}}$ of the relative velocity $w_{\mathrm{r,SS}}(r)$ is

\begin{equation}
\label{eq:GaussianPDF}
    p_{\mathrm{r,SS}}(w_{\mathrm{r,SS}}) = \frac{1}{\sqrt{2 \pi} \sigma_{\mathrm{SS}}} \exp{\left(-\frac{w_{\mathrm{r,SS}}^2}{2 \sigma_{\mathrm{SS}}^2}\right)}.
\end{equation}

\noindent
The standard deviation is determined by the mean relative velocity at $r$ as $\sigma_{\mathrm{SS}} = \sqrt{\pi/2} \cdot \langle |w_{\mathrm{r,SS}}| \rangle (r)$. It should be noted that, the relative velocity distribution is actually \emph{non-Gaussian} \citep{SundaramJFM1997, WangJFM2000, IrelandJFM2016a}. However, a Gaussian distribution is sufficient for the first-order assumption, which has already been applied in previous studies \citep{PanJFM2010, LuPOF2015}.

We then evaluate the RDF of charged pairs. For neutral small-small pairs, all particle pairs with an inward relative velocity ($w_{\mathrm{r,SS}}<0$) could approach each other. In comparison, when they are identically charged, only those SS pairs whose inward relative velocity exceeds a critical velocity ($w_{\mathrm{r,SS}}<-w_{\mathrm{C,SS}}$) are able to get close. By balancing the Coulomb energy barrier and the relative kinetic energy, the critical velocity can be given as

\begin{equation}
    w_{\mathrm{C,SS}}=\left( \frac{q^2}{2 \pi \varepsilon_0 r m_{\mathrm{SS}}}\right)^{1/2}.
\end{equation}

\noindent
The critical velocity for LL or SL pairs can be given by replacing $m_{\mathrm{SS}}=m_{\mathrm{S}}/2$ with $m_{\mathrm{LL}}=m_{\mathrm{L}}/2$ or $m_{\mathrm{SL}}=m_{\mathrm{S}}m_{\mathrm{L}}/(m_{\mathrm{S}}+m_{\mathrm{L}})$. Note that the critical velocity is derived from the interaction between a pair of particles, which  could reasonably describe the major effect of Coulomb force in the current dilution suspension. If the particle concentration becomes higher, the particle number density needs to be included for a more accurate prediction \citep{BoutsikakisJFM2022, BoutsikakisJCP2023}. By assuming a sharp cut-off at $w=-w_{\mathrm{C,SS}}$, \citet{LuPOF2015} related the RDF of charged SS pairs $g_{\mathrm{SS}}(r|q)$ to that of neutral SS pairs $g_{\mathrm{SS}}(r|q=0)$ as

\begin{equation}
\label{eq:gr_model_SS}
    g_{\mathrm{SS}}(r|q) = g_{\mathrm{SS}}(r|q=0) \cdot \frac{\int_{-\infty}^{-w_{\mathrm{C,SS}}} p_{\mathrm{r,SS}}(w) \mathrm{d}w}{\int_{-\infty}^{0} p_{\mathrm{r,SS}}(w) \mathrm{d}w}=g_{\mathrm{r,SS}}(r|q=0) \cdot \left[1-\mathrm{erf}\left(\frac{w_{\mathrm{C,SS}}}{\sqrt{2} \sigma}\right)\right],
\end{equation}

\noindent
where $\mathrm{erf}$ is the error function. However, as the relative velocity magnitude $w_{\mathrm{r}}$ becomes smaller than $w_{\mathrm{C,SS}}$, the corresponding flux contribution does not drop sharply to zero. Instead, a smooth transition would be expected, so the ratio of the charged RDF to the neutral RDF should be written as

\begin{equation}
\label{eq:Convolution}
    \frac{g_{\mathrm{SS}}(r|q)}{g_{\mathrm{SS}}(r|q=0)} =  \frac{\int_{-\infty}^{0} \Theta_{\mathrm{SS}}(w) p_{\mathrm{r,SS}}(w) \mathrm{d}w}{\int_{-\infty}^{0} p_{\mathrm{r,SS}}(w) \mathrm{d}w}.
\end{equation}

\noindent
Here, $\Theta_{\mathrm{SS}}(w)$ is the electrical kernel that gradually transients from one to zero when the magnitude of $w_{\mathrm{r}}$ drops below $w_{\mathrm{C,SS}}$. The proportion of charged particles that could overcome the Coulomb repulsion is then obtained from the convolution of $\Theta_{\mathrm{SS}}$ and $p_{\mathrm{r,SS}}$, which is the numerator of the right-hand term in Eq.\ref{eq:Convolution}. To account for the smooth transition without adding significant complexity, we still adopt the sharp cut-off assumption, but the \emph{effective} cut-off velocity $\beta_{\mathrm{SS}}w_{\mathrm{C,SS}}$ is used. Here, $\beta_{\mathrm{SS}} \sim O(1)$ is the correction factor that ensures an accurate prediction as: 

\begin{equation}
    \int_{-\infty}^{0} \Theta_{\mathrm{SS}}(w) p_{\mathrm{r,SS}}(w) \mathrm{d}w \approx \int_{-\infty}^{-\beta_{\mathrm{SS}}w_{\mathrm{C,SS}}} p_{\mathrm{r,SS}}(w) \mathrm{d}w.
\end{equation}

\noindent
The difference between the effective cut-off and the actual electrical kernel will be further discussed in Sec.\ref{sec:3}. Consequently, Eq.\ref{eq:gr_model_SS} becomes

\begin{equation}
\label{eq:gr_model_SS_final}
    g_{\mathrm{SS}}(r|q) = g_{\mathrm{SS}}(r|q=0) \cdot \left[1-\mathrm{erf}\left(\frac{\beta_{\mathrm{SS}} \cdot w_{\mathrm{C,SS}}}{\sqrt{2} \sigma}\right)\right].
\end{equation}

\noindent
To obtain the mean relative velocity between charged SS pairs, we start by computing the mean relative velocity in the velocity interval $[-\infty,-\beta_{\mathrm{SS}} w_{\mathrm{C,SS}}]$ between neutral SS pairs:

\begin{equation}
\label{eq:Cutoff_Velocity}
\begin{split}
    \langle |w_{\mathrm{r,SS}}^{\prime}| \rangle (r|q=0) & = \frac{-\int_{-\infty}^{-\beta_{\mathrm{SS}} w_{\mathrm{C,SS}}} w \cdot p_{\mathrm{r,SS}}(w) \mathrm{d}w}{\int_{-\infty}^{-\beta_{\mathrm{SS}} w_{\mathrm{C,SS}}} p_{\mathrm{r,SS}}(w) \mathrm{d}w} \\
    & = \langle |w_{\mathrm{r,SS}}| \rangle (r|q=0) \cdot \exp{\left( \frac{-\beta_{\mathrm{SS}}^2 w_{\mathrm{C,SS}}^2}{2 \sigma^2} \right)/\left[1-\mathrm{erf}\left(\frac{\beta_{\mathrm{SS}} \cdot w_{\mathrm{C,SS}}}{\sqrt{2} \sigma}\right)\right]}
\end{split}
\end{equation}

\noindent
Then the mean relative velocity of charged SS pairs can be obtained by subtracting the Coulomb potential energy from the mean relative kinetic energy in the velocity interval $[-\infty,-\beta_{\mathrm{SS}} w_{\mathrm{C,SS}}]$

\begin{equation}
\label{eq:EnergyBalance}
    \frac{1}{2}m_{\mathrm{SS}} {\langle |w_{\mathrm{r,SS}}| \rangle^2 (r|q)} = \frac{1}{2}m_{\mathrm{SS}} {\langle |w_{\mathrm{r,SS}}^{\prime}| \rangle^2 (r|q=0)} - \frac{\beta_{\mathrm{SS}}^2 q^2}{4 \pi \varepsilon_0 r}.
\end{equation}

\noindent
Here, the correction factor $\beta_{\mathrm{SS}}$ is also added to the last term on the right-hand side to account for the smooth transition. Taking Eq.\ref{eq:Cutoff_Velocity} into Eq.\ref{eq:EnergyBalance} then yields

\begin{equation}
\label{eq:wr_model_SS}
    \langle |w_{\mathrm{r,SS}}| \rangle (r|q) = \langle |w_{\mathrm{r,SS}}| \rangle (r|q=0) \cdot \left[ \frac{\exp{\left(-\frac{\beta_{\mathrm{SS}}^2 w_{\mathrm{C,SS}}^2 }{\sigma^2}\right)}}{ \left[1-\mathrm{erf} \left(\frac{\beta_{\mathrm{SS}}w_{\mathrm{C,SS}}}{\sqrt{2}\sigma}\right) \right]^2} - \frac{\beta_{\mathrm{SS}}^2 q^2}{ 2 \pi \varepsilon_0 m_{\mathrm{SS}} \cdot \langle |w_{\mathrm{r,SS}}| \rangle^2 (r|q=0) \cdot r } \right]^{1/2}
\end{equation}

Finally, by multiplying Eqs.\ref{eq:gr_model_SS} and \ref{eq:wr_model_SS} and taking into the definition of $\mathrm{Ct}_0$ (Eq.\ref{eq:Ct_0}), the flux ratio can be given as

\begin{equation}
\label{eq:FluxRatioModel_SS}
\begin{split}
    \frac{\Phi_{\mathrm{SS}}(q)}{\Phi_{\mathrm{SS}}(q=0)} & = \frac{g_{\mathrm{SS}}(r|q) \cdot \langle |w_{\mathrm{r,SS}}| \rangle (r|q)}{g_{\mathrm{SS}}(r|q=0) \cdot \langle |w_{\mathrm{r,SS}}| \rangle (r|q=0)} \\
    & = \left\{ \exp{\left(-\frac{2 \beta_{\mathrm{SS}}^2 \mathrm{Ct}_0}{\pi}\right)} - \beta_{\mathrm{SS}}^2 \mathrm{Ct}_0 [1-\mathrm{erf}(\beta_{\mathrm{SS}} \sqrt{\mathrm{Ct}_0 / \pi})]^2 \right\}^{1/2}.
    \end{split}
\end{equation}

The flux ratios for LL pairs and SL pairs can be derived in a similar way and are written as

\begin{subequations}
\label{eq:FluxRatioModel_LL_SL}
    \begin{equation}
    \frac{\Phi_{\mathrm{LL}}(q)}{\Phi_{\mathrm{LL}}(q=0)} = \left\{ \exp{\left(-\frac{2 \beta_{\mathrm{LL}}^2 \mathrm{Ct}_0}{\pi}\right)} - \beta_{\mathrm{LL}}^2 \mathrm{Ct}_0 [1-\mathrm{erf}(\beta_{\mathrm{LL}} \sqrt{\mathrm{Ct}_0 / \pi})]^2 \right\}^{1/2},
    \end{equation}

    \begin{equation}
    \frac{\Phi_{\mathrm{SL}}(q)}{\Phi_{\mathrm{SL}}(q=0)} = \left\{ \exp{\left(-\frac{2 \beta_{\mathrm{SL}}^2 \mathrm{Ct}_0}{\pi}\right)} + \beta_{\mathrm{SL}}^2 \mathrm{Ct}_0 [1+\mathrm{erf}(\beta_{\mathrm{SL}} \sqrt{\mathrm{Ct}_0 / \pi})]^2 \right\}^{1/2}.
\end{equation}
\end{subequations}

Eqs.\ref{eq:FluxRatioModel_SS} and \ref{eq:FluxRatioModel_LL_SL} are then fitted as blue lines in Fig.\ref{fig:5}, which show good agreement with the simulation results. The fitted correction factors are $\beta_{\mathrm{SS}}=0.193$, $\beta_{\mathrm{LL}}=0.739$, and $\beta_{\mathrm{SL}}=0.5415$, satisfying the assumption that the correction factors are of the order of unity. The predictions with the fixed $\beta=1$ are also shown as brown dash-dotted lines, which correspond to the original model that assumes the sharp cut-off occurs at the critical velocity $w_{\mathrm{C,SS/LL/SL}}$. Although the trends are similar, models with the fixed $\beta=1$ underestimate the critical $\mathrm{Ct}_0$ at which the transition occurs. Moreover, the proposed model underestimates the SS flux when $\mathrm{Ct}_0 \approx 10^2$ in Fig.\ref{fig:5}(a). To find the origin of this discrepancy, the predicted $g_{ij}(r)$ using Eq. \ref{eq:gr_model_SS_final} and the predicted $\langle |w_{\mathrm{r,SS}}| \rangle (r)$ using Eq. \ref{eq:wr_model_SS} are also plotted as grey dashed lines in Fig. \ref{fig:2} and Fig. \ref{fig:3}, respectively. It can been seen that, the predicted RDFs are comparable to the simulation results, so the discrepancy of $\Phi_{\mathrm{SS}}$ at $\mathrm{Ct}_0 \approx 10^2$ mainly comes from the underestimation of the mean relative velocity at $r \sim \eta$ in Fig. 3(a). Since the PDF of $w_{\mathrm{r}}$ is assumed Gaussian in our model, the influence of intermittency is not considered. As a result, the proportional of particle pairs with large relative velocity is significantly underestimated.

\subsection{Particle flux density for charged particles}
\label{sec:3}

In Sec.\ref{sec:2}, the mean Coulomb-turbulence parameter $\mathrm{Ct}_0$ is defined using the mean relative velocity $\langle |w_{\mathrm{r},ij}| \rangle$, which compares the importance of Coulomb force to the mean relative motion caused by turbulence. However, it has been known that the relative velocity between particle pairs may vary within a wide range, and the mean Coulomb-turbulence parameter $\mathrm{Ct}_0$ may not be sufficient to fully reveal the physics. 

In this section it would be of our interest to examine the impacts of Coulomb force on particle pairs with different relative velocities. For different kinds of particle pairs, the particle flux $\Phi_{ij}$ in the relative velocity space is expanded as

\begin{equation}
\label{eq:FluxExpansion}
    \Phi_{ij} = \int_{-\infty}^{\infty} \phi_{ij}(w_{\mathrm{r}}) \mathrm{d} w_{\mathrm{r}}.
\end{equation}

\noindent
Here, $\phi_{ij}$ is the density of particle flux within each relative velocity interval, which measures the contribution to the total particle flux $\Phi_{ij}$ by particle pairs whose relative velocity is within the interval $w_{\mathrm{r}} \pm \Delta w_{\mathrm{r}}/2$. By comparing Eqs. \ref{eq:FluxExpansion} and \ref{eq:FullFlux}, $\phi_{ij}$ can be given as

\begin{equation}
\label{eq:FluxDensity}
    \phi_{ij}=\frac{1}{2} g_{ij} \cdot |w_{\mathrm{r},ij}|p_{ij}(w_{\mathrm{r},ij}).
\end{equation}

\begin{figure}
    \centering
    \includegraphics[width=13.5cm]{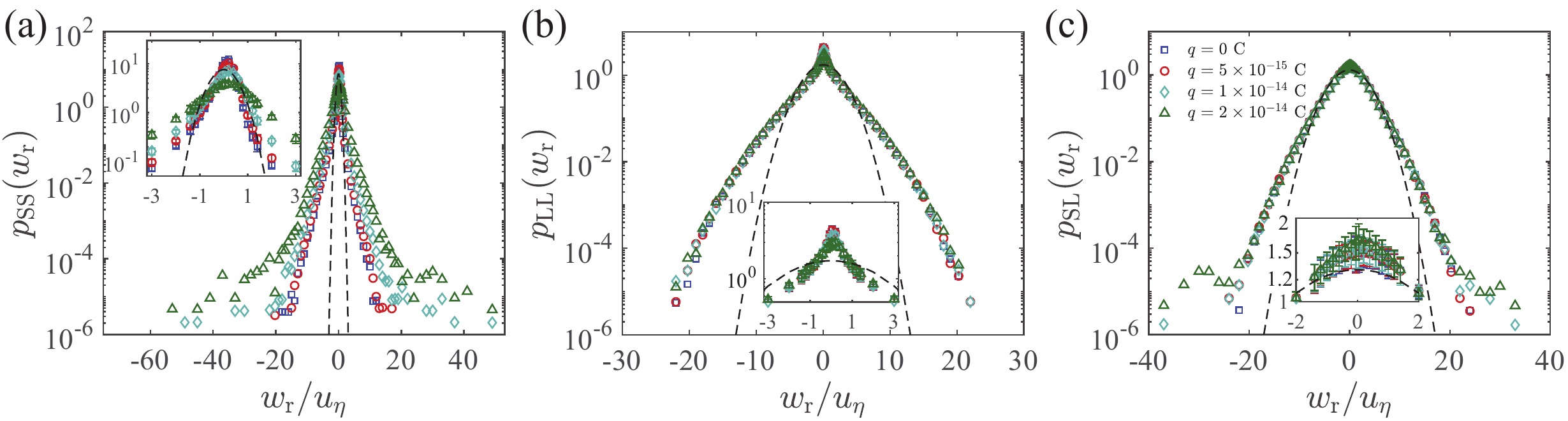}
    \caption{PDFs of the relative velocity between (a) small-small, (b) large-large, and (c) small-large pairs at $r \in [ 1.15\eta, 2.5\eta )$. Zoom-in views around $w_{\mathrm{r}}=0$ are shown in the insets. Black dashed lines denote the Gaussian distributions defined by Eq. \ref{eq:GaussianPDF} for different kinds of particle pairs.}
    \label{fig:6}
\end{figure}

We start with the effect of Coulomb force on the PDFs of relative velocity $p_{ij}(w_{\mathrm{r},ij})$, because the distribution of $\phi_{ij}$ in Eq.\ref{eq:FluxDensity} is strongly dependent on $p_{ij}(w_{\mathrm{r},ij})$. Fig. \ref{fig:6}(a) illustrates PDFs of $w_{\mathrm{r,SS}}$ at the separation distance interval $r \in [1.15\eta, 2.5\eta)$. The Gaussian distribution defined by Eq. \ref{eq:GaussianPDF} is also shown as the black dashed line. By comparing the neutral PDFs and the Gaussian curves, it is clear that the Gaussian assumption serves as a reasonable approximation at $|w_{\mathrm{r}}|<3u_\eta$, but significantly underestimates the probability of large $|w_{\mathrm{r}}|$. Therefore, the Gaussian assumption is only suitable for modeling low-order effects, not for higher-order statistics. Besides, as will be discussed below, since the Gaussian curve is symmetric, it is not able to capture the asymmetry of the relative velocity between approaching and departing pairs. 

For SS pairs, compared to the neutral PDF, charged PDFs become wider as $q$ increases, indicating a lower/higher probability of finding particle pairs with low/high relative velocity within the separation interval. This is consistent with the increase of the mean relative velocity with $q$ in Fig.\ref{fig:3}(a). If we look at the symmetry of $p_{\mathrm{SS}}$, the neutral PDF is found negatively-skewed. This asymmetry is attributed to: (1) the fluid velocity derivative is negatively-skewed, and the relative motion between particle pairs with $St=1$ is partially coupled with local flow and thus exhibits a similar feature \citep{VanAttaPOF1980, WangJFM2000}; (2) the asymmetry of particle path history gives rise to larger relative velocity between approaching pairs than departing pairs \citep{BraggNJP2014}. However, the asymmetry of $p_{\mathrm{SS}}$ curves are significantly reduced once particles are charged (see Table \ref{tab:Skewness}), which results from the symmetric nature of the Coulomb force. The magnitude of Coulomb force relies only on the interparticle distance $r$, so the approaching or departing SS pairs experience the same amount of repulsion as long as $r$ is the same, which makes the approaching-then-departing process more symmetric. Therefore, introducing Coulomb force could effectively increase the standard deviation but reduce the skewness of $w_{\mathrm{r,SS}}$.

\begin{table}
  \begin{center}
\def~{\hphantom{0}}
  \begin{tabular}{cccc}
    $q (\times 10^{-14} \ \mathrm{C})$ & $\langle w_{\mathrm{r,SS}}^3 \rangle / [\langle w_{\mathrm{r,SS}}^2 \rangle]^{3/2}$ & $\langle w_{\mathrm{r,LL}}^3 \rangle / [\langle w_{\mathrm{r,LL}}^2 \rangle]^{3/2}$ & $\langle w_{\mathrm{r,SL}}^3 \rangle / [\langle w_{\mathrm{r,SL}}^2 \rangle]^{3/2}$\\
     $0$ & $-2.144$ & $-0.074$ & $-0.081$\\
     $0.5$ & $-1.817$ & $-0.079$ & $-0.093$\\
     $1$ & $-1.066$ & $-0.073$ & $-0.089$\\
     $2$ & $-0.500$ & $-0.075$ & $-0.114$\\
    \end{tabular}
  \caption{Normalized skewness of the radial relative velocity $w_{\mathrm{r}}$ at $r \in [ 1.15\eta, 2.5\eta )$.}
  \label{tab:Skewness}
  \end{center}
\end{table}

For LL pairs, since Coulomb repulsion is only able to repel pairs with small relative velocity, $p_{\mathrm{LL}}$ drops slightly as $w_{\mathrm{r}}$ approaches zero, but remains almost unchanged at larger $w_{\mathrm{r}}$. As for $p_{\mathrm{SL}}$, the difference between neutral and charged case is insignificant, which again demonstrates that the velocity difference between particles of different sizes mainly comes from the differential inertia effect, while the influence of Coulomb force is limited. Besides, the movement of large particles with $St=10$ has very weak couplings with local flow fields, so the skewness of $p_{\mathrm{LL}}$ and $p_{\mathrm{SL}}$ in Table \ref{tab:Skewness} are negligible.

\begin{figure}
    \centering
    \includegraphics[width=10cm]{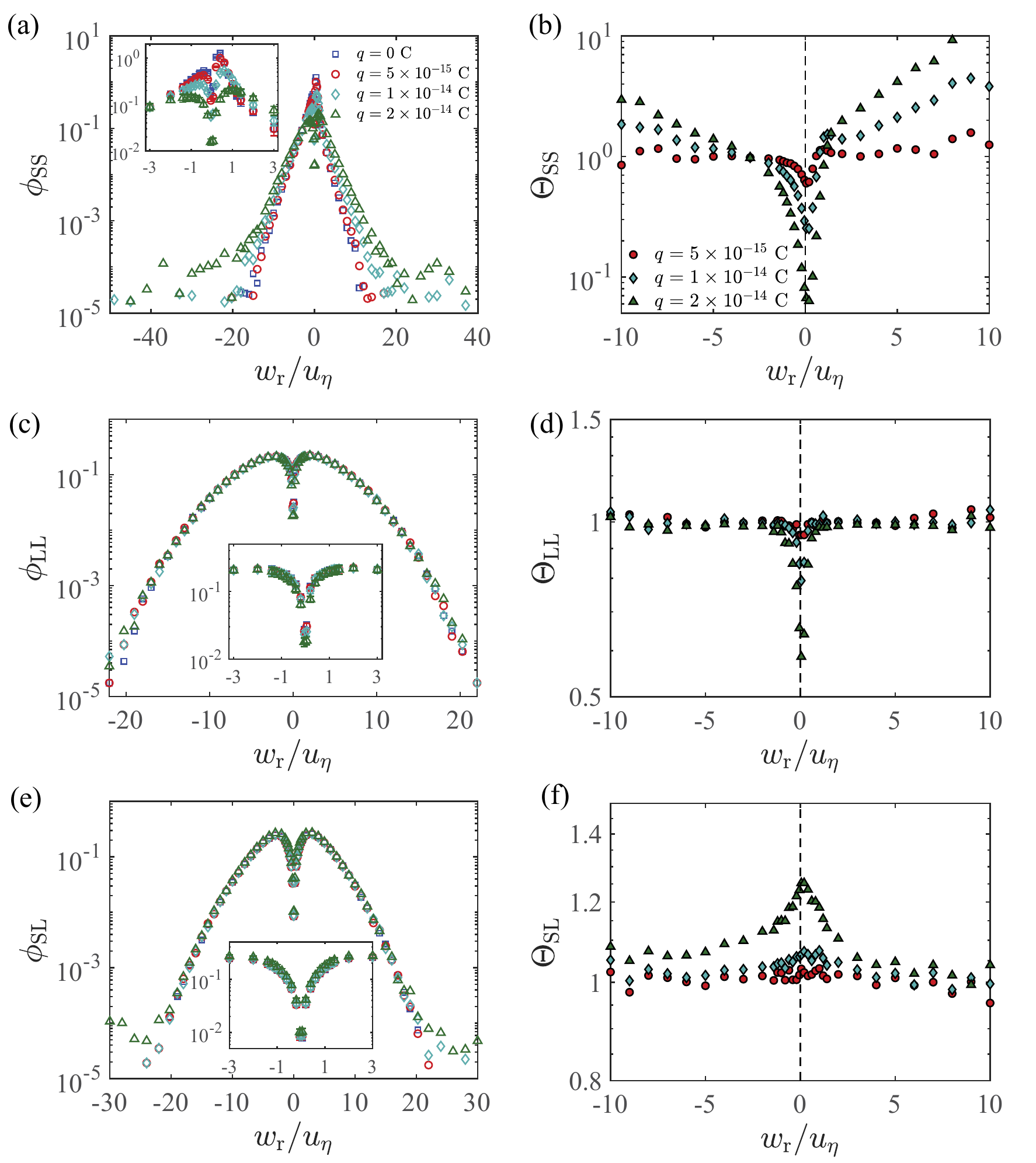}
    \caption{Particle flux density in the velocity space, $\phi(w_{\mathrm{r}})$, between (a) small-small, (c) large-large, and (e) small-large pairs at $r \in [ 1.15\eta, 2.5\eta )$. Zoom-in views around $w_{\mathrm{r}}=0$ are shown in the insets. (b), (d), (f): Dependence of Coulomb kernel $\Theta_{ij}=\phi_{ij}(q)/\phi_{ij}(q=0)$ on $w_{\mathrm{r},ij}$ corresponding to (a), (c), (e), respectively.}
    \label{fig:7}
\end{figure}

We now turn to the distribution of flux density $\phi_{ij}$ in the velocity space. Fig.\ref{fig:7}(a),(c),(e) plot the flux densities between SS, LL, and SL pairs, respectively. Different from the unimodal PDFs of the relative velocity shown in Fig.\ref{fig:6}, $\phi_{ij}$ are bimodal because the maximum flux density is reached when the product $|w_{\mathrm{r},ij}| \cdot p_{ij}(w_{\mathrm{r},ij})$ is the largest. Therefore, it is the particle pairs with the intermediate relative velocity that contributes the most to the overall particle flux.

To better describe the impacts of Coulomb force, we define the Coulomb kernel $\Theta_{ij}$ as the ratios of the charged flux densities to their neutral values, i.e., $\Theta_{ij}(q) = \phi_{ij}(q)/\phi_{ij}(q=0)$, which are displayed in the right panel of Fig.\ref{fig:7}. As shown in Fig.\ref{fig:7}(b), the value of $\Theta_{\mathrm{SS}}$ varies by more than one order of magnitude, indicating that the influence of Coulomb force changes significantly as $w_{\mathrm{r}}$ changes. When $|w_{\mathrm{r}}|$ is small, $\Theta_{\mathrm{SS}}$ decreases drastically as $q$ increases because Coulomb repulsion dominates. In addition, $\Theta_{\mathrm{SS}}$ is found to be asymmetric. As discussed above, neutral particles with $St=1$ tend to separate at a lower relative velocity compared with the approaching pairs. When they are charged, particle pairs originally separating at low speeds will be accelerated and pushed apart by Coulomb repulsion, which effectively shifts the high flux density from the small positive $w_{\mathrm{r}}$ to a larger positive $w_{\mathrm{r}}$ in the velocity space. Consequently, $\Theta_{\mathrm{SS}}$ experiences a more significant decrease at $w_{\mathrm{r}}$ slightly greater than zero, but quickly exceeds one when $w_{\mathrm{r}} \approx 2 u_\eta$. Moreover, $\Theta_{\mathrm{SS}}$ becomes larger than one at large $|w_{\mathrm{r}}|$ for both approaching and departing pairs. The explanation is that, with the increase of $q$, small particles are more likely to get attracted towards the large ones and follow their motions. Since the relative velocity between LL pairs is generally much larger, this effect could increase the SS flux density at large $|w_{\mathrm{r}}|$. However, the contribution of the increased flux at large positive $|w_{\mathrm{r}}|$ is negligible compared to the decrease at small $|w_{\mathrm{r}}|$ (see Fig.\ref{fig:7}(a)), so the major effect of Coulomb force is to reduce the SS flux through the small-small repulsive force. 

The kernel $\Theta_{\mathrm{LL}}$ between LL pairs (Fig.\ref{fig:7}(d)) also drops quickly at small $|w_{\mathrm{r}}|$, but recovers to unity when $|w_{\mathrm{r}}| \geq 2u_\eta$. Besides, because of the limited effects of local fluid velocity gradient mentioned above, the approaching and departing processes are more symmetric for neutral LL pairs, and  adding Coulomb force still retains this symmetry. As for $\Theta_{\mathrm{SL}}$ shown in Fig.\ref{fig:7}(f), the change is still most significant at small $w_{\mathrm{r}}=0$. However, different from the kernels of SS/LL pairs where the impact of the same-sign repulsion is only obvious within a narrow interval (i.e., $|w_{\mathrm{r}}|/u_\eta \leq 2$), the opposite-sign attraction seems to increase $\Theta_{\mathrm{SL}}$ in a much wider range of $w_{\mathrm{r}}$. For instance, $\Theta_{\mathrm{SL}}$ is increased even at $w_{\mathrm{r}}/u_\eta=-10$ for $q=2\times10^{-14} \ \mathrm{C}$ in Fig.\ref{fig:7}(f). As discussed in Sec.\ref{sec:2}, the main effect of the opposite-sign attraction is to enhance small-large clustering. Then particles of different sizes, though staying close to each other, will develop a large relative velocity as a result of their different responses to local fluctuations, which leads to the increase of $\Theta_{\mathrm{SL}}$ in a wide range of $w_{\mathrm{r}}$.

The discussion above has shown that, the influence of Coulomb force is different as the particle relative velocity $w_{\mathrm{r},ij}$ changes. Therefore, instead of using the mean relative velocity $\langle |w_{\mathrm{r},ij}| \rangle$ (Eq.\ref{eq:Ct_0}), we adopt the median relative velocity $\overline{w}_{\mathrm{r},ij}$ in each $w_{\mathrm{r}}$ bin to estimate the relative kinetic energy. For a given separation distance $r$ and a certain relative velocity bin with the median $\overline{w}_{\mathrm{r},ij}$, the extended Coulomb-turbulence parameter is given as

\begin{equation}
\label{eq:Ct_ext}
    \mathrm{Ct} \equiv \frac{q_i q_j}{2 \pi \varepsilon_0 r m_{ij} \overline{w}_{\mathrm{r},ij}^2}.
\end{equation}

We then examine the dependence of the electrical kernel $\Theta_{ij}$ on the extended Coulomb-turbulence parameter $\mathrm{Ct}$. For the data points $(w_{\mathrm{r},ij},\Theta_{ij})$ shown in the right panel of Fig.\ref{fig:7}, the corresponding values of $\mathrm{Ct}$ are calculated using Eq.\ref{eq:Ct_ext}, and the results are re-plotted as points $(\mathrm{Ct}, \Theta_{ij})$. Note that to ensure meaningful statistics, only data points satisfying certain criterion are considered and the reasons are as follows:

\hspace*{\fill}

\hangindent 2em
(1) In addition to the separation interval $r \in [1.15\eta,2.5\eta)$ shown in Fig.\ref{fig:7}, data from two more separation intervals, i.e., $[0.85\eta,1.15\eta)$ and $[2.5\eta,3.5\eta)$, are also added. At these separation distances, the effect of electrostatic force is already significant, while the number of samples is large enough for statistics.

\hangindent 2em
(2) We only consider pairs with their relative velocity in a certain range to avoid using the more scattered data points at large $w_{\mathrm{r}}$. For SS pairs, the relative velocity range is $[-5 u_\eta, 5 u_\eta]$, while for LL/SL pairs the range considered is $[-10 u_\eta, 10 u_\eta]$. The particle flux within the above ranges contribute to at least $97.1 \% / 95.6\% / 96.3\%$ of the total SS/LL/SL particle flux in the neutral case, which reflects the major change of $\Phi_{ij}$ in each case.

\hangindent 2em
(3) As shown in Fig.\ref{fig:7}(b), $\Theta_{\mathrm{SS}}$ is not symmetric about $w_{\mathrm{r}}=0$. When particles are departing, Coulomb repulsion will redistribute $\phi_{\mathrm{SS}}$ in the velocity space, which may distort the $\Theta_{\mathrm{SS}}-\mathrm{Ct}$ relationship. We therefore only use the data from approaching pairs ($w_{\mathrm{r}}<0$) for later analysis. Since the outward flux is balanced by the inward flux in the steady state, the total flux could still be evaluated as $\Phi_{\mathrm{SS}}(q)=2\int_{-\infty}^{0} \Theta_{\mathrm{SS}}(\mathrm{Ct}) \phi_{\mathrm{SS}}(w_{\mathrm{r,SS}}) \mathrm{d} w_{\mathrm{r,SS}}$.

\begin{figure}
    \centering
    \includegraphics[width=13.5cm]{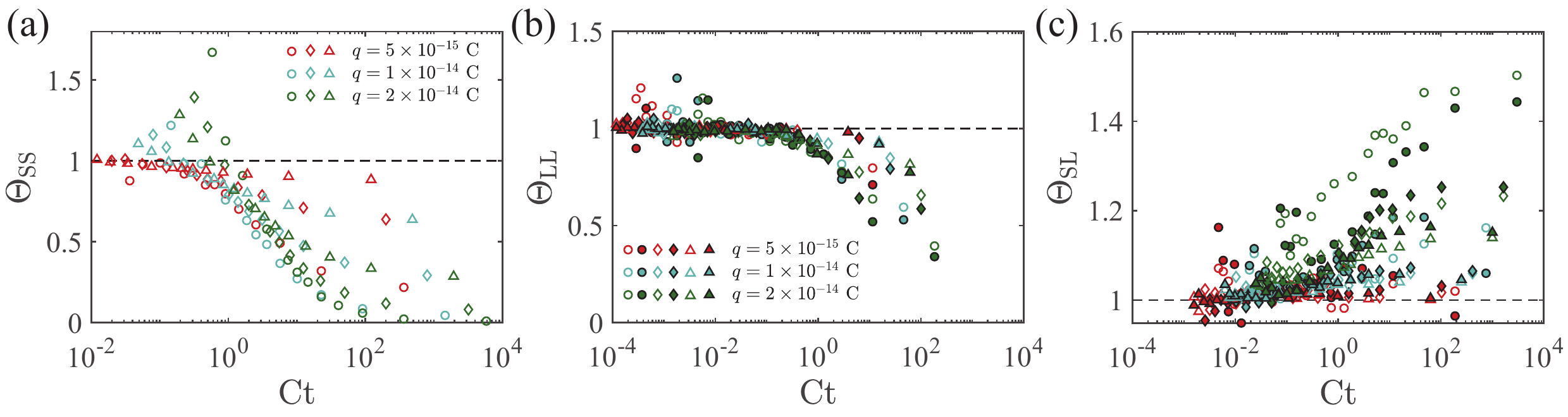}
    \caption{Dependence of the electrical kernel $\Theta_{ij}$ on the extended Coulomb-turbulence parameter $\mathrm{Ct}$ for (a) small-small, (b) large-large, and (c) small-large particle pairs. Different shapes represent data points from $r \in [0.85\eta,1.15\eta)$ (circles: $\circ$), $r \in [1.15\eta,2.5\eta)$ (diamonds: $\diamond$), and $r \in [2.5\eta,3.5\eta)$ (triangles: $\triangle$), respectively. Open/filled symbols denote data points for approaching/departing pairs.}
    \label{fig:8}
\end{figure}

\hspace*{\fill}

Fig.\ref{fig:8} plots the dependence of $\Theta_{ij}$ on $\mathrm{Ct}$ for different particle pairs. Despite different particle charge $q$ and separation $r$, the data points for both SS (Fig.\ref{fig:8}(a)) and LL (Fig.\ref{fig:8}(b)) pairs show a similar trend. When $\mathrm{Ct}$ is small, Coulomb force is weak compared with particle-turbulence interaction, so $\Theta_{\mathrm{SS}}$ and $\Theta_{\mathrm{LL}}$ stay close to one. When $\mathrm{Ct}$ becomes larger than one, the same-sign repulsion starts to significantly reduce the corresponding particle flux density, and a decrease of $\Theta$ is observed. In Sec.\ref{sec:2}, it has been assumed that a sharp cut-off occurs at the effective velocity $\beta_{\mathrm{SS}} w_{\mathrm{C,SS}}$ or $\beta_{\mathrm{LL}} w_{\mathrm{C,LL}}$ for SS or LL pairs. However, the transition of $\Theta_{\mathrm{SS}}$ and $\Theta_{\mathrm{LL}}$ in Fig.\ref{fig:8}(a) turn out to be smooth. Thus, the sharp cut-off can be understood as the first-order approximation of the influence of Coulomb repulsion, which can be written as

\begin{equation}
\label{eq:Theta_model}
    \Theta_{ij}(\mathrm{Ct}) = 1-H(\mathrm{Ct}_{\mathrm{crit},ij}),
\end{equation}

\noindent
with $H(\cdot)$ being the Heaviside step function. The critical value of $\mathrm{Ct}$ describing where $\Theta_{ij}$ steps down can be related to the fitted correction factors for the effective cut-off velocity as

\begin{equation}
\label{eq:Ct_crit}
\mathrm{Ct}_{\mathrm{crit},ij}=1/\beta_{ij}^2.
\end{equation}

\noindent
By taking into the fitted results ($\beta_{\mathrm{SS}}=0.193$ and $\beta_{\mathrm{LL}}=0.739$) from Section \ref{sec:2}, the critical value can be given as $\mathrm{Ct}_{\mathrm{crit,SS}} = 26.85 $ and $\mathrm{Ct}_{\mathrm{crit,LL}}=1.83$, respectively.

However, different from $\Theta_{\mathrm{SS}}$ and $\Theta_{\mathrm{LL}}$, there is no clear dependence of $\Theta_{\mathrm{SL}}$ on $\mathrm{Ct}$ in Fig.\ref{fig:8}(c). The reason is as follows. The opposite-sign Coulomb force will attract SL pairs with a low departing velocity together and promote clustering. However, although these SL pairs have a low relative velocity at first, they will develop a much larger relative velocity over time. As a result, the relative kinetic energy $E_{\mathrm{Kin}}$ becomes independent of the electrical potential energy $E_{\mathrm{Coul}}$, and no clear relationship can be found between $\Theta_{\mathrm{SL}}$ and $\phi_{\mathrm{SL}}$.

\section{Conclusions}
In this study, we investigate the effects of charge segregation on the dynamics of tribocharged bidispersed particles in homogeneous isotropic turbulence by means of DNS. Using radial distribution function $g_{ij}(r)$, we show that Coulomb repulsion/attraction significantly inhibits/promotes particle clustering within a short range, while the clustering at a large separation distance is still determined by particle-turbulence interaction. For same-sign particles, Coulomb repulsion repels approaching pair with low relative velocity, giving rise to the increase of mean relative velocity $\langle |w_{\mathrm{r},ij}| \rangle$ as the separation distance $r$ decreases. In comparison, the relative velocity between different-size particles is almost unchanged for all $q$, which suggests that the differential inertia effect contributes predominantly to $\langle w_{\mathrm{r,SL}} \rangle$.

By defining the particle flux $\Phi_{ij}$ as the number of particles crossing the collision sphere per area per unit time, we are able to quantify the particle collision frequency without a prescribed collision diameter $R_{\mathrm{C}}$. As expected, the Coulomb repulsion/attraction is found to reduce/increase the total particle flux $\Phi_{ij}$ when particles are close. When plotted as a function of the mean Coulomb-turbulence parameter $\mathrm{Ct}_0$ that measures the relative importance of electrostatic potential energy to the mean relative kinetic energy, the particle flux ratio $\Phi_{ij}(q)/\Phi_{ij}(q=0)$ is shown to follow a similar trend. Specifically, by assuming that the PDF of relative velocity follows a Gaussian distribution, the particle flux can be well modelled by a function of $\mathrm{Ct}_0$.

The total particle flux $\Phi_{ij}$ is then expanded in the relative velocity space as the flux density $\phi_{ij}$, and the relative change $\Theta_{ij}=\phi_{ij}(q)/\phi_{ij}(q=0)$ (also termed the Coulomb kernel) in each relative interval exhibits a significant difference. Finally, the extended Coulomb-turbulence parameter $\mathrm{Ct}$ is defined using the median $\overline{w}_{\mathrm{r},ij}$ in each relative velocity bin, which better describes the competition between Coulomb force and the turbulence effect. Then for same-size particle pairs, a clear relationship is found between $\Theta_{ij}$ and $\mathrm{Ct}$. And the smooth transition of $\Theta_{ij}$ is observed near the critical value $\mathrm{Ct}_{\mathrm{crit},ij}=1/\beta_{ij}^2$. For SL pairs, however, the relative velocity will grow larger because of the predominant differential inertia effect, so $\Theta_{\mathrm{SL}}$ shows no clear dependence on $w_{\mathrm{r,SL}}$ (and therefore on $\mathrm{Ct}$).

\section*{Acknowledgements}
This work was supported by an Early Stage Innovation grant from NASA's Space Technology Research Grants Program under Grant NO. 80NSSC21K0222. This work was also partially supported by the Office of Naval Research (ONR) under Grant NO. N00014-21-1-2620.

\section*{Declaration of interests}
The authors report no conflict of interest.

\appendix
\section{Validation of the electrostatic calculation}
\label{appA}

To validate the electrostatic computation introduced in Section \ref{sec:Partciel_Motion}, we compute the Coulomb force acting on $N_{\mathrm{p}}$ particles in the 3D periodic box using both (1) FMM incorporated with periodic image boxes and (2) the standard Ewald summation. For the charge-neutral system in this work, the exact Coulomb force acting on particle $i$ can be computed by Ewald summation \citep{DesernoJChemPhys1998} as:  

\begin{equation}
    \boldsymbol{F}^{\mathrm{E,Ewald}}_i = \boldsymbol{F}^{(r)}_i + \boldsymbol{F}^{(k)}_i + \boldsymbol{F}^{(d)}_i,
\end{equation}

\noindent
where the contribution from the real (physical) space $\boldsymbol{F}^{\mathrm{(r)}}_i$, the Fourier (wavenumber) space $\boldsymbol{F}^{\mathrm{(k)}}_i$, and the dipole correction $\boldsymbol{F}^{\mathrm{(d)}}_i$ are given as

\begin{subequations}
    \begin{equation}
    \label{eq:EwaldReal}
        \boldsymbol{F}^{\mathrm{(r)}}_i = \frac{q_i}{4 \pi \varepsilon_0} \sum_j q_j \sum_{\boldsymbol{m} \in \mathbb{Z}^3}^{\prime} \left( \frac{2 \alpha}{\sqrt{\pi}} \mathrm{exp} \left( -\alpha^2 |\boldsymbol{r}_{ij} + \boldsymbol{m}L|^2 \right) + \frac{\mathrm{erfc} \left( \alpha |\boldsymbol{r}_{ij} +\boldsymbol{m}L| \right)}{|\boldsymbol{r}_{ij} +\boldsymbol{m}L|} \right) \frac{\boldsymbol{r}_{ij} +\boldsymbol{m}L}{|\boldsymbol{r}_{ij} +\boldsymbol{m}L|^2},
    \end{equation}

    \begin{equation}
    \label{eq:Fourier}
        \boldsymbol{F}^{\mathrm{(k)}}_i = \frac{q_i}{4 \pi \varepsilon_0 L^3} \sum_j q_j \sum_{\boldsymbol{k} \neq \boldsymbol{0}} \frac{4 \pi \boldsymbol{k}}{k^2} \mathrm{exp} \left( -\frac{k^2}{4 \alpha^2} \right) \sin{(\boldsymbol{k} \cdot \boldsymbol{r}_{ij})},
    \end{equation}

    \begin{equation}
    \label{eq:EwaldDipole}
        \boldsymbol{F}^{\mathrm{(d)}}_i = -\frac{q_i}{\varepsilon_0 (1+2 \varepsilon^{\prime})L^3} \sum_j q_j \boldsymbol{x}_j.
    \end{equation}
\end{subequations}

\noindent
Here, $\alpha$ is the Ewald parameter, $\mathrm{erfc}$ is the complimentary error function, and $\varepsilon^{\prime}=1$ is the relative dielectric constant of surrounding medium. 

Since Ewald summation is computationally expensive, a smaller-scale system with $N_{\mathrm{p}}=2000$ oppositely charged particles is used in validation cases. In each case, ten parallel computations with different particle locations are performed to ensure reliable statistics. Table \ref{tab:EwaldParameters} lists parameters used in Ewald summation. The dimensionless product $\alpha r_{\mathrm{C}}$ equals $\pi$ to ensure high accuracy in both real and Fourier spaces. The cut-off radius($r_{\mathrm{C}}$)/wavenumber($k_{\mathrm{C}}$) in the real/Fourier space is then determined by $r_{\mathrm{C}}=(\alpha r_{\mathrm{C}}) L/\pi^{1/2}N_{\mathrm{p}}^{1/6}$ and $k_{\mathrm{C}} = 1.8 (\alpha r_{\mathrm{C}})^2/r_{\mathrm{C}}$ to balance the computation cost of $\boldsymbol{F}^{\mathrm{(r)}}_i$ and $\boldsymbol{F}^{\mathrm{(k)}}_i$ \citep{FinchamMolSimulation1994}.

\begin{table}
  \begin{center}
\def~{\hphantom{0}}
  \begin{tabular}{lll}
    Parameters & Symbols & Values\\
    Domain size & $L$ & $2 \pi$\\
    Particle number & $N_{\mathrm{p}}$ & $2000$\\
    Particle charge & $q$ & $\pm 1$\\
    Accuracy parameter & $\alpha r_{\mathrm{C}}$ & $\pi$\\
    Cut-off distance in real space & $r_{\mathrm{C}}$ & $\pi$ \\
    Cut-off wavenumber in Fourier space & $k_{\mathrm{C}}$ & $6$\\
    Error in real space & $\varepsilon^{\mathrm{(r)}}$ & $1.65 \times 10^{-5}$\\
    Error in Fourier space & $\varepsilon^{\mathrm{(k)}}$ & $2.06 \times 10^{-5}$\\
    \end{tabular}
  \caption{Dimensionless parameters in Ewald summation.}
  \label{tab:EwaldParameters}
  \end{center}
\end{table}

When performing FMM computation, the layer number $N_{\mathrm{per}}$ is varied from $0$ to $4$ to show the influence of adding image domains. The relative error of FMM compared to Ewald summation is given as

\begin{equation}
    \epsilon^{\mathrm{FMM}} = \frac{|\boldsymbol{F}^{\mathrm{E,FMM}}-\boldsymbol{F}^{\mathrm{E,Ewald}}|}{|\boldsymbol{F}^{\mathrm{E,Ewald}}|} =\frac{[ \sum_{i=1}^{N_\mathrm{p}} (\boldsymbol{F}_i^{\mathrm{E,FMM}}-\boldsymbol{F}_i^{\mathrm{E,Ewald}})^2 /N_{\mathrm{p}}]^{1/2}}{[ \sum_{i=1}^{N_\mathrm{p}} (\boldsymbol{F}_i^{\mathrm{E,Ewald}})^2 /N_{\mathrm{p}}]^{1/2}},
\end{equation}

\noindent
where the norm of force is defined as the root mean square of the force components following \citet{DesernoJChemPhys1998}. The dependence of $\epsilon^{\mathrm{FMM}}$ on $N_{\mathrm{
per}}$ is shown in Table \ref{tab:ESValidation}. The relative error is significant ($\epsilon^{\mathrm{FMM}} >10 \%$) if periodicity is no considered. After adding image domains, the relative error drops significantly and almost saturates after $N_{\mathrm{per}} \geq 2$. Hence,  $N_{\mathrm{per}} = 2$ is chosen to guarantee sufficient accuracy ($\epsilon^{\mathrm{FMM}}=0.17\%$) while avoiding expensive computation cost. 

\begin{table}
  \begin{center}
\def~{\hphantom{0}}
  \begin{tabular}{lccccc}
    $N_{\mathrm{per}}$ & $0$ & $1$ & $2$ & $3$ & $4$ \\
    $\epsilon^{\mathrm{FMM}}$ & $1.12 \times 10^{-1}$ & $2.13 \times 10^{-3}$ & $1.70 \times 10^{-3}$ & $1.65 \times 10^{-3}$ & $1.63 \times 10^{-3}$ \\
    \end{tabular}
  \caption{Relative errors of FMM computation compared to Ewald summation.}
  \label{tab:ESValidation}
  \end{center}
\end{table}

\bibliographystyle{jfm}
\bibliography{jfm-instructions}

\end{document}